\documentclass[nofootinbib,preprint,onecolumn,showpacs,preprintnumbers,amsmath,amssymb]{revtex4-1}

\usepackage{graphicx}
\usepackage{dcolumn}
\usepackage{bm}
\usepackage[T1]{fontenc}
\usepackage{amsmath}
\usepackage{amssymb}
\usepackage{amsfonts}
\usepackage{mathrsfs}

\usepackage{float}
\restylefloat{figure}
\usepackage{epstopdf}
\newcommand{\bea} {\begin{equation}\ba{lcl}}
\newcommand{\eea} {\ea\end{equation}}
\def\lf{\left}
\def\ri{\right}
\def\Jc{{\cal J}}
\def\Nc{{\cal N}}

\def\s{\hat s}
\def\t{\hat t}
\def\u{\hat u}
\def\ksl{\not{\hbox{\kern-2.3pt $k$}}}
\newcommand{\comment}[1]{}

\newcommand{\be}{\begin{equation}}
\newcommand{\ee}{\end{equation}}

\newcommand{\req}[1]{(\ref{#1})}

\def\fc#1#2{\frac{#1}{#2}}
\def\h{\frac{1}{2}}

\newcommand{\ba}{\begin{array}}
\newcommand{\ea}{\end{array}}

\newcommand{\bc}{\begin{center}}
\newcommand{\ec}{\end{center}}
\newcommand{\ds}{\displaystyle}
\newcommand{\p} {\partial}
\newcommand{\Tr}{{\rm Tr}}

\def\rng{\rangle}
\def\Mc{{\cal M}}

\def\ov{\overline}
\def\th{\theta}

\def\s{{\hat s}}

\newcommand{\beq}{\begin{equation}}
\newcommand{\eeq}{\end{equation}}

\def\eps{\varepsilon}

\newcommand{\vecb}{\left(\begin{array}{c}}
\newcommand{\vece}{\end{array}\right)}
\newcommand{\ccb}{\left(\begin{array}{cc}}
\newcommand{\cce}{\end{array}\right)}
\newcommand{\cccb}{\left(\begin{array}{ccc}}
\newcommand{\ccce}{\end{array}\right)}
\newcommand{\ccccb}{\left(\begin{array}{cccc}}
\newcommand{\cccce}{\end{array}\right)}
\newcommand{\cccccb}{\left(\begin{array}{ccccc}}
\newcommand{\ccccce}{\end{array}\right)}

\newcommand{\pa}{\partial}

\newcommand{\al}{\alpha}

\newcommand{\de}{\delta}

\newcommand{\si}{\sigma}

\newcommand{\Ga}{\Gamma}

\newcommand{\Si}{\Sigma}

\newcommand{\Om}{\Omega}

\newcommand{\te}{\textrm}
\newcommand{\eq}{ \ \ = \ \ }

\newcommand{\co}{\ , \ \ \ \ \ \ }

\newcommand{\dd}{d}

\newcommand{\dal}{\dot{\alpha}}
\newcommand{\dbe}{\dot{\beta}}

\newcommand{\hpl}{{+}{\textstyle\frac{1}{2}}}
\newcommand{\hmi}{{-}{\textstyle\frac{1}{2}}}
\newcommand{\tpl}{{+}{\textstyle\frac{3}{2}}}
\newcommand{\tmi}{{-}{\textstyle\frac{3}{2}}}
\def \ap{\alpha'}

\def\fc#1#2{{\frac{#1}{#2}}}
\def\lf{\left}\def\ri{\right}

\def\h{\fc{1}{2}}
\def\req#1{(\ref{#1})}

\def\ie{{\it i.e.\ }}
\def\p{\partial}

\def\Oc{{\cal O}}
\def\Tr{{\rm Tr}}

\def\ds{\displaystyle}
\def\cf{{\it c.f.\ }}
\def\ie{{\it i.e.\ }}

\def\ra{\rightarrow}

\begin{document}

\title{Direct Production of Lightest Regge Resonances}

\author{\vskip1cm W.Z. Feng$^{a,b}$, D. L\"ust$^{b,c}$,  O. Schlotterer$^b$,
S. Stieberger$^b$, T.R. Taylor$^{a,b,c}$}

\affiliation{\vskip1cm
$^a$  Department of Physics, Northeastern University,
Boston, MA 02115, USA\\
$^b$ Max--Planck--Institut f\"ur Physik
Werner--Heisenberg--Institut,
80805 M\"unchen, Germany\\
$^c$ Arnold--Sommerfeld--Center for Theoretical Physics,
Ludwig--Maximilians--Universit\"at, 80333 M\"unchen, Germany \vskip4cm}

\preprint{MPP--2010--93}
\preprint{LMU--ASC 60/10}


\begin{abstract}
We discuss direct production of Regge excitations in the collisions of massless
four--dimensional superstring states, focusing on the first excited level of
open strings
ending on D--branes extending into higher dimensions. We construct covariant
vertex
operators and identify
``universal'' Regge states with the internal parts either trivial or determined
by the world--sheet SCFT
describing superstrings propagating on an arbitrary Calabi--Yau manifold.
We evaluate the amplitudes
involving  one such massive state and up to three massless ones and express them
in the helicity basis. The most important phenomenological applications of our
results
are in the context of low--mass string (and large extra dimensions)
scenarios in which excited string states are expected
to be produced at the LHC as soon as the string mass threshold is reached in
the center--of--mass energies of the colliding partons.
In order to facilitate the use of
partonic cross sections, we evaluate them and tabulate for all production
processes:
gluon fusion, quark absorbing a gluon, quark--antiquark annihilation and
quark-quark scattering.
\end{abstract}

\maketitle
\tableofcontents
\break

\section{Introduction}
The most direct way to exhibit the existence of fundamental strings is by
discovering the effects of their vibrations. The particles that appear as
the quanta of  oscillating string modes are called Regge excitations and have
squared masses quantized in the units of $M^2=1/\al'$, where $\al'$ is the
Regge slope. Direct detection
of such vibrating string modes is possible at the LHC, provided that $M$ is in
the range of few TeVs. For a survey of low-mass superstring phenomenology and
early
references, see
Refs. \cite{Antoniadis:1997zg,
Antoniadis:1998ig,Cullen:2000ef,Burikham:2004su}.
At first, one would see Regge excitations indirectly, in the excess of photons
\cite{Anchordoqui:2007da,Anchordoqui:2008ac}, jets
\cite{Anchordoqui:2008di,Anchordoqui:2009mm}, heavy quarks \cite{Dong:2010jt}
and leptons due to the resonant enhancement of their production rates.
As emphasized in
Refs. \cite{Anchordoqui:2007da,Anchordoqui:2008ac,Lust:2008qc,Lust:2009pz}, some
of
the most interesting signals of low-mass superstring theory are completely
model-independent, {\it i.e}.\ they do not depend  on the compactification
space or on the SUSY breaking mechanism, hence
by studying such processes we can avoid the landscape problem\footnote{
The effects of Regge resonances and Kaluza--Klein (KK) gravitons are also discussed in
Refs. \cite{Accomando:1999sj,Chemtob:2008cb,Dudas:1999gz,Chialva:2005gt}.}.

Once the mass threshold $M$ is crossed in the center--of--mass energies of the
colliding partons, one would also see free Regge states produced directly, in
association with jets, photons and other particles. In this paper, we discuss
direct production of lightest Regge particles, {\it i.e}.\ the quanta of
fundamental string harmonics with masses equal $M$.

The basic property of Regge states is that they populate linear trajectories,
with the slope  $\al'$,
correlating their masses and spins. At the $n$--th mass level,
\begin{equation}\label{reggemasses}
M_n^2\ = \ n\ M^2\ \ \ ,\ \ \ n \in {\bf N}\ ,
\end{equation}
the spins range from 0 to
$n{+}1$. However, the spectrum of Regge excitations are highly model-dependent.
{}For example, in the toroidal compactifications of a single ten-dimensional
$D9$-brane
one encounters 128 bosons and 128 fermions at the $n{=}1$ level. Most of these
particles are tied to
${\cal N}{=}4$ supersymmetry of toroidal compactifications, however some of them
are universal to all Calabi-Yau manifolds. No other Regge states than the
universal ones can be observed as resonances in gluon-gluon scattering or in the
processes involving only gluons and one pair of external fermions originating
from D--brane intersections\footnote{For a detailed account on D--brane
constructions,
see Ref. \cite{Blumenhagen:2006ci}.}.
In Ref. \cite{Anchordoqui:2008hi}, the spin content and multiplicities of the
universal part of the first level have been disentangled by factorizing the
four-point amplitudes.
In this paper, we construct the covariant vertex operators creating universal
particles. We include not only trivial internal parts but also the operators
appearing in the world-sheet superconformal field theory (SCFT)
of a superstring propagating on an arbitrary Calabi-Yau manifold (CYM). They
create particles present in all compactifications preserving at least ${\cal
N}{=}1$
SUSY and can decay into (or be created by the annihilation of) two gluons and/or
gluinos. Furthermore, we construct the vertex operators for the lightest excited
quarks, assuming that massless quarks originate from
superstrings stretching between D-branes intersecting in the internal space.

With the vertex operators at hand, it is not too difficult to compute the
scattering amplitudes involving one  massive particle and two or three massless
ones. The amplitudes relevant to the direct production of string resonances at
the LHC are $p_1p_2\to p_3R$, where $p~(=g,q,\bar q)$ are partons and $R$ is a
massive string state.

The paper is organized as follows. In Section II we discuss some general factorization
properties of string amplitudes.
In Section III, we discuss the first massive level of open superstrings
originating from a stack of $N$ D-branes extending into higher dimensions.
Among the universal excitations of gauge bosons, we find
one spin 2 particle, one vector and two complex scalars, with only spin 2 and a
single scalar coupled to massless gauge bosons directly at the disk level.
On the other hand, quarks exist in the excited spin 3/2 and 1/2 states. We
construct the corresponding vertex operators. In Section IV, the wave functions
describing massive particles with spins up to $J{=}2$
are written in the van der Waerden (helicity) representation.
In Section V, we compute all amplitudes involving one of the universal Regge
excitations and up to three massless partons. These amplitudes acquire a very
simple form in the helicity basis, which also reveals certain selection rules
similar (and related) to the vanishing of ``all-plus'' amplitudes at the zero
mass level \cite{Stieberger:2007jv}.
In Section VI, we square the appropriately crossed amplitudes for  $p_1p_2\to
p_3R$, average
over initial helicities and colors and sum over the colors and spin directions
of the outgoing particles.
In order to facilitate phenomenological applications of the partonic cross
sections, we tabulate squared amplitudes according to the production processes:
gluon fusion, gluon-quark absorption, quark-antiquark annihilation and
quark-quark scattering.

\section{Parton amplitudes and factorization on massive poles}

The most direct way to see string effects is to  measure the open string
oscillator excitations, the so--called Regge
modes. There are infinitely many open string Regge (SR) modes
and their mass--squares are multiples of the string scale \req{reggemasses}.
To each level $n$ a set of states $|J,n\rng$ belongs.
The latter are classified by their spin $J$
with the highest possible spin $n+1$ limited by the oscillator number $n$.
Each Standard Model (SM) particle is accompanied by an infinite number of these
massive string
excitations, {\it i.e}.\, the latter carry SM quantum numbers and thus
may be produced by $pp$--collisions.

SR excitations may appear in resonance channels of SM processes
or may be directly produced as external states.
While the first effect has been extensively studied in
\cite{Cullen:2000ef,Lust:2008qc,Lust:2009pz}
the latter effect will be discussed in this work.
A first look at the couplings of massless SM particles to massive SR states is
made
by considering the factorization of higher--point amplitudes involving massless
external states.
In what follows we shall discuss this factorization on general grounds.
We consider scattering amplitudes involving massless SM model
open string fields $\Phi_i$ as external particles.
These amplitudes are described\footnote{There may be
additional resonance channels
due to the exchange of KK and winding states, as it is the case for amplitudes
involving at least four quarks or leptons.}
by the exchange of the light (massless) SM fields
and the tower of infinite many higher SR excitations.

Due to the extended nature of strings the  string amplitudes
are generically non--trivial functions of $\ap$ in addition to the
usual dependence on the kinematic invariants and degrees of freedom of the
external states.
In the effective field theory description this $\ap$--dependence gives rise to a
series of infinite many resonance channels
due to Regge excitations and new contact interactions involving massless
SM fields and massive SR states.
As a consequence of unitarity an $N$--point tree--level string
amplitude\footnote{Disk amplitudes $\Mc(\Phi_1,\ldots,\Phi_N)$
involving $N$ open string states $\Phi_i$
as external states decompose into a sum over all
possible orderings $\rho$ of the corresponding vertex operators $V_{\Phi_i}$
along the boundary of the disk
\be\label{BASIC}
\Mc(\Phi_1,\ldots,\Phi_N)=\sum_{\rho\in S_N}\Mc_\rho(\Phi_1,\ldots,\Phi_N)
=\sum_{\rho\in S_N}\Tr(T^{a_{1_\rho}}\ldots T^{a_{N_\rho}})\
A(1_\rho,\ldots,N_\rho)\ ,
\ee
with $i_\rho=\rho(i)$ and the partial ordered amplitudes
$A(1_\rho,\ldots,N_\rho)$.
Furthermore, $T^{a_i}$ is the Chan--Paton factor accounting for the
gauge degrees of freedom of the two ends of the $i$--th open string.}
$\Mc(\Phi_1,\ldots,\Phi_N)$ can be written as an infinite
sum over exchanges of (massive) intermediate
string states $|J,n\rangle$ coupling to
$N_1$ and $N_2$ external massless string states,
with $N_1+N_2=N$. For each level this
pole expansion gives rise to (new) $N_1+1$-- and $N_2+1$--point
couplings between the massive string states $|J,n\rangle$ and the
$N_1$ and $N_2$ external massless string states, respectively.

In the following we illustrate this at the four--gluon amplitude, {\it i.e}.\
$\Phi_i=g_i$ and $N_1=N_2=2$.
The latter gives rise to an infinite series of three--point
couplings involving two massless gluons and  massive string states $|J,n\rangle$.
The general expression for the four--gluon amplitude in $D$ space--time dimensions is
\bea\label{FinalIV}
\ds{\Mc(g_1,g_2,g_3,g_4)}&=&\ds{2\ g_{YM}^2\
K_4(\eps_1,k_1;\eps_2,k_2;\eps_3,k_3;\eps_4,k_4)}\\[5mm]
&\times&\ds{\lf\{\
T^{a_1a_2a_3a_4}\ \fc{B(\s,\u)}{\t}+T^{a_2a_3a_1a_4}\ \fc{B(\t,\u)}{\s}
+T^{a_3a_1a_2a_4}\ \fc{B(\s,\t)}{\u}\ \ri\}\ ,}
\eea
with the Euler Beta function
\be\label{Euler}
B(x,y)=\fc{\Gamma(x)\ \Gamma(y)}{\Gamma(x+y)}\ ,
\ee
the kinematic factor \cite{Green:1981xx,Schwarz:1982jn}
\bea\label{t8}
&&K_4(\eps_1,k_1;\eps_2,k_2;\eps_3,k_3;\eps_4,k_4)=
\t\u\ (\eps_1\eps_2)\ (\eps_3\eps_4)
+\s\t\ (\eps_1\eps_4)\ (\eps_2\eps_3)+\s\u\ (\eps_1\eps_3)\ (\eps_2\eps_4)\\
&&\hskip0.45cm+  \s\ \lf[\ (\eps_1\eps_3)(\eps_2k_3)(\eps_4k_1)+
(\eps_1\eps_4)(\eps_2k_4)(\eps_3k_1) +
(\eps_2\eps_3)(\eps_1k_3)(\eps_4k_2)+(\eps_2\eps_4)(\eps_1k_4)(\eps_3k_2)\
\ri]\\
&&\hskip0.45cm+  \t\ \lf[\
(\eps_1\eps_2)(\eps_3k_2)(\eps_4k_1)+(\eps_1\eps_4)(\eps_2k_1)(\eps_3k_4) +
(\eps_2\eps_3)(\eps_1k_2)(\eps_4k_3)+(\eps_3\eps_4)(\eps_1k_4)(\eps_2k_3)\
\ri]\\
&&\hskip0.45cm+  \u\ \lf[\
(\eps_1\eps_2)(\eps_3k_1)(\eps_4k_2)+(\eps_1\eps_3)(\eps_2k_1)(\eps_4k_3) +
(\eps_2\eps_4)(\eps_1k_2)(\eps_3k_4)+(\eps_3\eps_4)(\eps_1k_3)(\eps_2k_4)\ \ri],
\eea
and the color factor:
\be\label{Tfact}
T^{a_1a_2a_3a_4} =
\Tr(T^{a_1}T^{a_2}T^{a_3}T^{a_4})+\Tr(T^{a_4}T^{a_3}T^{a_2}T^{a_1})\ .
\ee
Above, $\eps_i$ are the polarization vectors and $k_i$ the external momenta
of the four gluons. Furthermore, we have the kinematic invariants
$\hat s=2\ap k_1k_2,\   \hat t=2\ap k_1k_3$ and $\hat u=2\ap k_1k_4$.

\begin{figure}[H]
\centering
\includegraphics[width=1.05\textwidth]{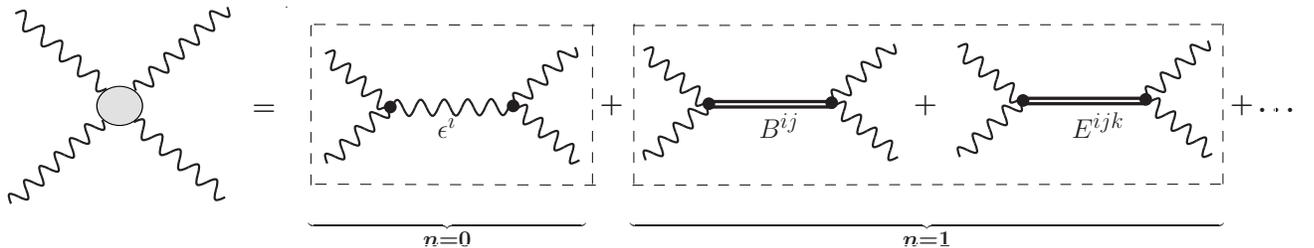}
\caption{\it Factorization of four--gluon amplitude
into pairs of two three--point couplings.}
\label{factorization}
\end{figure}

In what follows we shall concentrate on the partial amplitude
$\Mc_{(1234)}$. According to the definition \req{BASIC} we have:
$\Mc_{(1234)}=\Tr(T^{a_1}T^{a_2}T^{a_3}T^{a_4})\ A(1,2,3,4)$.
With (c.f. Ref. \cite{Lust:2008qc})
\be\label{basic}
\fc{B(\hat s,\hat u)}{\hat t}=\fc{1}{\hat t\hat u}\ \fc{\Gamma(\hat s)\
\Gamma(1+\hat u)}{\Gamma(\hat s+\hat u)}
=\sum_{n=0}^\infty\ \ \fc{\gamma(n)}{\hat s+n}\ ,
\ee
and
\be\label{DEG}
\gamma(n)={1\over n!}\ \fc{\Gamma(\u+ n)}{\Gamma(\u+1)}=
\fc{1}{n!}\ \fc{1}{\u}\ \prod\limits_{j=1}^n(\u-1+j)
\ee
the amplitude \req{FinalIV} can be written as an infinite sum over
$s$--channel poles at the masses \req{reggemasses}
of the SR excitations:
\be\label{FinalIVa}
\Mc_{(1234)}(g_1,g_2,g_3,g_4)=2\ g_{YM}^2\ \Tr(T^{a_1}T^{a_2}T^{a_3}T^{a_4})\
K_4(\eps_1,k_1;\eps_2,k_2;\eps_3,k_3;\eps_4,k_4)\
\sum_{n=0}^\infty\ \fc{\gamma(n)}{\s+n}\ .
\ee
In \req{FinalIVa} to each residue at $\s=-n$
a class of three--point couplings of two massless and one massive SR
state $|J,n\rangle$ of a specific spin $J$ is associated, c.f.
Fig. \ref{factorization}.

\noindent
From \req{FinalIVa} these three--point couplings
are determined by the product $2\ g_{YM}^2\ \gamma(n)\ K_4$.
To cast the residua of \req{FinalIVa} into suitable form
non--trivial factorization properties of the kinematic factor \req{t8} have to
hold.

We shall now evaluate for the amplitude \req{FinalIVa}
the contribution to the residue of the pole in $\s$ at $\s=-n$,
with $n=0,1,\ldots$.
At the level $n=0$ only a massless gluon
with polarization $\eps^i$ and spin $J=1$ is exchanged.
Hence, we obtain the following residue at $\s=0$
\bea\label{nice1}
\ds{{\rm Res}_{\s=0\atop \u=-\t}\ \Mc_{(1234)}(g_1,g_2,g_3,g_4)}&=&
\ds{2\ g_{YM}^2\ \Tr(T^{a_1}T^{a_2}T^{a_3}T^{a_4})}\\
&\times&\ds{\gamma(0)
\lf.K_4(\eps_1,k_1;\eps_2,k_2;\eps_3,k_3;\eps_4,k_4)\ri|_{\s=0\atop \u=-\t}}\\
&=&\ds{\sum_{\eps(k)}K_{3,0}(\eps_1,k_1;\eps_2,k_2;\eps,k)\
K_{3,0}(\eps_3,k_3;\eps_4,k_4;\eps,-k)\ ,}
\eea
with the YM three--vertex:
\bea\label{YM3}
\ds{K_{3,0}(\eps_1,k_1;\eps_2,k_2;\eps_3,k_3)}&=&\ds{
g_{YM}\ \Tr(T^{a_1}[T^{a_2},T^{a_3}])}\\
&\times&\ds{\lf\{\ (\eps_1\eps_2)\ (\eps_3k_1)+(\eps_1\eps_3)\ (\eps_2k_3)+
(\eps_2\eps_3)\ (\eps_1k_2)\ \ri\}\ .}
\eea
Furthermore we have applied the completeness relations
\be\label{completeness}
\sum_a\Tr(T^{a_1}T^{a_2}T^a)\ \Tr(T^aT^{a_3}T^{a_4})=\h\
\Tr(T^{a_1}T^{a_2}T^{a_3}T^{a_4})\ ,
\ee
with the sum over the Chan--Paton wavefunction of the
intermediate state.

At the $n=1$ level  exchanges of a spin $J=2$ state $B^{ij}$ and a $J=0$
state $E^{ijk}$ occur (\cf \cite{Anchordoqui:2008hi} for $D=4$).
For the amplitude \req{FinalIVa}
we obtain the following residue at $\s=-1$
\bea\label{nice2}
\ds{{\rm Res}_{\s=-1\atop \u=1-\t}\ \Mc_{(1234)}(g_1,g_2,g_3,g_4)}&=&
\ds{2\ g_{YM}^2\ \Tr(T^{a_1}T^{a_2}T^{a_3}T^{a_4})} \\
&\times&\ds{\lf.\gamma(1)\ K_4(\eps_1,k_1;\eps_2,k_2;\eps_3,k_3;\eps_4,k_4)
\ri|_{\s=-1\atop \u=1-\t}}\\
&=&\ds{\sum_{E(k)}K_{3,1}(\eps_1,k_1;\eps_2,k_2;E,k)\
K_{3,1}(\eps_3,k_3;\eps_4,k_4;E,-k)}\\
&+&\ds{\sum_{B(k)}K_{3,2}(\eps_1,k_1;\eps_2,k_2;B,k)\
K_{3,2}(\eps_3,k_3;\eps_4,k_4;B,-k)\ ,}
\eea
with the two three--point vertices (in $D$ space--time dimensions)
\begin{eqnarray}
\ds{K_{3,1}(\eps_1,k_1;\eps_2,k_2;E,k)}&=&
\ds{6\ g_{YM}\ \lf\{\Tr(T^{a_1}T^{a_2}T^{a_3})+\Tr(T^{a_2}T^{a_1}T^{a_3})\ri\}\
E^{ijk}\ \eps_{1i}\ \eps_{2j}\ k_{1k}\ ,}\label{Massive1}\\[5mm]
\ds{K_{3,2}(\eps_1,k_1;\eps_2,k_2;B,k)}&=&
\ds{g_{YM}\ \lf\{\Tr(T^{a_1}T^{a_2}T^{a_3})+\Tr(T^{a_2}T^{a_1}T^{a_3})\ri\}}
\label{Massive2}\\
&\times&\ds{B^{ij}\ \lf\{\ (k_1k_2)\ \eps_{1i}\eps_{2j}-
k_{1i}\eps_{2j}\ (\eps_1k_2)-k_{2i}\eps_{1j}\ (\eps_2k_1)+
k_{1i}k_{2j}\ (\eps_1\eps_2)\ \ri\}\ ,}\nonumber
\end{eqnarray}
involving two massless gluons and one massive string state $E^{ijk}$
and $B^{ij}$, respectively.
In \req{nice2} the second equality follows by applying results from
\cite{Yasuda:1987pu,Xiao:2005yn,Xiao:2004vs}.

\section{\label{spectrum}The first massive level}\noindent

The first massive level of $D=10$  superstring, at the mass equal to the
fundamental string scale $M$,
consists of
the quanta of fundamental harmonics.
It is well known that it contains
128 bosonic degrees of freedom from the Neveu-Schwarz (NS) sector and 128
fermionic degrees of freedom from the Ramond (R) sector.
These particles have the same Chan--Paton gauge charges as gauge bosons and
gauginos because
they all originate from open
strings ending on D9-branes in $D=10$ spacetime. Hence it is appropriate to
call them massive  (or excited) ``gluons'' and ``gluinos''. In addition, in
specific models, there are similar particles associated to stacks of lower
dimensional D-branes, with a reduced number of degrees of freedom.
Furthermore, quarks, leptons and other MSSM fields arising from
D-brane intersections, also appear in excited states at this level.

In Refs. \cite{Lust:2008qc,Lust:2009pz} we stressed the
universality (model--independence) of certain scattering amplitudes with an
arbitrary number of external gluons and at most two fermions, at the leading
disk order of string perturbations theory. This property allows testing the low
string mass scenario at the LHC, in dijet mass distributions
\cite{Anchordoqui:2008di} and other physical quantities
\cite{Anchordoqui:2009mm}, in a completely model--independent manner, thus
nullifying the notorious landscape problem. Our present goal is to identify the
string excitations that can be produced on-shell, in parton
collisions at center-of-mass energies above the string threshold mass $M$.
Since the universal disk amplitudes do not depend on the extent of
four-dimensional supersymmetry, it is clear that not all first level particles
can be produced in such collisions.  To some extent, this goal has been
accomplished in Ref. \cite{Anchordoqui:2008hi}, where the relevant amplitudes
have been factorized at the $s=M^2$ poles, exhibiting spin 2 and spin 0
resonances propagating in two-gluon channels {\em etc}. In this Section we go
one step farther and identify the vertex operators of universal resonances. We
will use them in the following Section to
compute the amplitudes involving one massive and three massless states,
which can be useful at the LHC for describing direct production of massive
string states with large transverse momentum. Since we want to avoid
discussing supersymmetry breaking already at this stage, we  begin by
truncating the first string level down to the universal
$\Nc=1$ supermultiplets in $D=4$.

For maximally supersymmetric, toroidal compactifications of $D=10$ superstring,
its excitations form supermultiplets of $\Nc=4$ supersymmetry. Before
discussing the first excited level, we recall
the vertices of massless particles, which arise from the zero modes and
include, in
the NS sector, the gluon $A$ and six real scalars $\phi^j, ~j=1,\dots,
6.$ In the R sector, we have four gauginos $\lambda^I, ~I=1,\dots, 4$. All
in all, these zero mode form one $\Nc=4$
gauge supermultiplet.
The NS sector vertices, in the $(-1)$-ghost picture, read:
\begin{eqnarray} \label{onepic}
V_{A^a}^{(-1)}(z,\epsilon,k) &=& g_A\ T^a\ e^{-\phi}\
\epsilon^{\mu}\, \psi_{\mu}\ e^{ikX}\ ,\cr
V_{\phi^{a,j}}^{(-1)}(z,k) &=& g_A\ T^a\ e^{-\phi}\ \Psi^{j}\
e^{ikX}.\end{eqnarray}
Here, $X,\psi,Z,\Psi$ are the fields of $\Nc=1$ worldsheet SCFT, with the Greek
indices associated to $D=4$ spacetime fields $X^\mu, \psi^\nu$ and the Latin
lower case labeling internal $D=6$ (e.g. $Z^i, \Psi^j$). $\phi$ is the scalar
bosonizing the superghost
system. {}For completeness, we write below the same vertices in the 0-ghost
picture:
\begin{eqnarray}\label{zeropic}
V_{A^a}^{(0)}(z,\epsilon,k) &=& \frac{g_A}{\sqrt{2\ap}} \ T^a\ \epsilon^\mu\
[\ i\partial X_\mu+2\ap \ k^\nu \psi_\nu\ \psi_\mu \ ]\ e^{ikX}\ ,\cr
V_{\phi^{a,j}}^{(0)}(z,k) &=& \frac{g_A}{\sqrt{2\ap}} \  T^a\
[\ i\partial Z^j+2\ap \,k^\nu \psi_\nu\ \Psi^j\ ] \ e^{ikX}\ .
\end{eqnarray}
The R sector vertices, in the $(-1/2)$-ghost picture, read:
\begin{eqnarray}\label{vertlam}
V_{\lambda^{a,I}}^{(-1/2)}(z,u,k)
&=&g_\lambda\ T^a\ e^{-\phi/2}\ u^{\sigma} S_{\sigma}\ \Si^I\ e^{ik X}\ ,\cr
V_{\bar\lambda^{a,I}}^{(-1/2)}(z,\bar u,k)
&=&g_\lambda\ T^a\ e^{-\phi/2}\ \bar u_{\dot\sigma} \bar S^{\dot\sigma}\
\ov\Si^I\
e^{ikX}.\end{eqnarray}
Here, $S$ and $\bar S$ are the left and right-handed $SU(2)$ spin fields,
respectively, while $\Si^I$ and $\ov\Si^I$ are the internal Ramond spin fields
discussed in the following Subsection.
The couplings are
\begin{equation}
g_A=(2\ap)^{1/2}\ g\quad,\quad g_\lambda=(2{\ap})^{1/2}{\ap}^{1/4}\ g\ ,
\end{equation}
where $g$ is the gauge coupling.
In the above definitions, $T^a$ are the Chan-Paton factors accounting for the
gauge
degrees of freedom of the two open string ends.

We will be also considering quarks and their first-level excitations. Assuming
that the $SU(3)$ color group is associated to stack $a$, the vertices of quarks
originating from strings attached at the other end to an intersecting stack $b$
are given by
\begin{eqnarray}
V_{q^{\al}_{\beta}}^{(-1/2)}(z,u,k) &=&  \sqrt{2} \  \alpha^{\prime 3/4} \,
e^{\phi_{10}/2} \  \bigl(T^{\al}_{\beta} \bigr)^{\beta_{1}}_{\al_{1}} \
u^{\sigma}   S_{\sigma} \  e^{-\phi/2} \   {\Xi}^{a \cap b}\ e^{ik
X}\ ,\label{vertq}\nonumber\\
V_{\bar{q}^{\beta}_{\al}}^{(-1/2)}(z,\bar{u},k) &=&  \sqrt{2} \  \alpha^{\prime
3/4} \  e^{\phi_{10}/2} \  \bigl(T^{\beta}_{\al} \bigr)^{\al_{1}}_{\beta_{1}} \
\bar{u}_{\dot\sigma}   \bar S^{\dot\sigma} \  e^{-\phi/2} \  \bar{\Xi}^{a \cap
b} \  e^{ik  X}\ ,\label{QUARKV}
\end{eqnarray}
where the Chan-Paton factors read
\begin{equation}(T^{\alpha_1}_{\,\beta_1})_{\alpha_2}^{\beta_2}=\delta^{\alpha_1
}_{\alpha_2}
\delta^{\beta_2}_{\beta_1}\end{equation}
and $\Xi^{a \cap b}$ are the fermionic boundary changing operators.
The  latter carry the internal degrees of freedom  of the Ramond sector
associated to the internal part of the SCFT.
Their two--point correlator is given by  \cite{Lust:2008qc}:
\be\label{INTCFT}
\langle \Xi^{a \cap b}(z_{1}) \, \bar{\Xi}^{a \cap b}(z_{2}) \rangle
=\fc{1}{(z_1-z_2)^{3/4}}\ .
\ee
In general, such particles are in the bi-fundamental representations of the
gauge groups associated to the two stacks.
Note that
the above vertices can be obtained from the gluino vertices (\ref{vertlam}) by
simply adjusting the Chan-Paton factors and replacing the internal spin
operators by the boundary changing operator. Thus up to such minor
modifications, the vertices of excited quarks can be be ``borrowed'' from the R
sector.

\subsection{\label{bosons}Bosons (NS sector)}\noindent
For maximally supersymmetric, toroidal compactifications of $D=10$ superstring,
NS and R sectors form one spin 2 ($J=2$) massive supermultiplet of $\Nc=4$
supersymmetry. The bosons form one symmetric tensor field $B_{mn}$ and one
completely antisymmetric tensor field $E_{mnp}$. Here, the indices $(m,n,p)$
label $D=10$. All these particles are in the adjoint representation of the
gauge group.
The corresponding vertices, in the $(-1)$-ghost picture, read
\cite{Koh:1987hm}:
\begin{equation}\label{verteb}
V_{N\! S,a}^{(-1)}(z,k) = \frac{g_A}{\sqrt{2\alpha'}}\ T^a\ e^{-\phi}(\,
E_{mnp}\,\psi^m\psi^n\psi^p \,
+
\, B_{mn}\, i\partial X^m\psi^n \, + \, H_m\partial\psi^m \,
)\,e^{ikX}\ ,
\end{equation}
where $H_m$ is an auxiliary vector field. Note that at this level, the on-shell
condition is $k^2=-\frac{1}{\ap}$. The constraints due to the requirement of BRS
invariance are:
\begin{eqnarray}\nonumber
~~~~k^mE_{mnp}&=&0\ ,\\
2\ap k^mB_{mn} + H_n &=&0\ ,\label{brstbos}\\
B^m_m+k^mH_m&=&0\ .\nonumber
\end{eqnarray}
In $D=10$ all 128 bosonic degrees of freedom can be accounted for by setting
$H=0$,
{\em i.e.} with a traceless, transverse $B$ and transverse $E$. However, this
auxiliary vector will be useful for discussing the spectrum of compactified
theory.

In order to find out which of these particles can be produced in purely gluonic
processes, we consider
the disk amplitude involving one massive boson  and an arbitrary number of
gluons.
Note that one of the gluon vertices must be in the $(-1)$-picture, see
(\ref{onepic}) while the rest in the 0-picture, see (\ref{zeropic}).
Thus in (\ref{verteb}), all world-sheet fermions  must be associated to $D=4$
coordinates: $\psi^m=\psi^{\mu}$ {\em etc}. Furthermore, at the disk level,
internal $\partial Z^i$ have no zero modes. We conclude that only
$E_{\mu\nu\rho}$ and $B_{\mu\nu}$ interact with gluons at the disk level.

In $D=4$, a massive particle described by a transverse three-form is equivalent
to a pseudoscalar with the wave function proportional to:
\begin{equation} E_{\mu\nu\rho}\sim
\epsilon_{\mu\nu\rho\lambda}\,k^{\lambda}.\end{equation}
A symmetric, transverse and {\em traceless} $B_{\mu\nu}$ gives rise to a spin 2
particle. However, this cannot be the end of the story. As pointed out in
Ref. \cite{Anchordoqui:2008hi}, pseudoscalar resonances give rise to four-gluon
scattering amplitudes with non-MHV helicity configurations, thus contradicting
the fact that such amplitudes are zero at the disk level
\cite{Stieberger:2007jv}. Thus we need another $J=0$ particle to cancel such
pseudoscalar contributions. The relevant scalar is described by the following
solution of the constraint (\ref{brstbos}):
\begin{equation} E=0\qquad,\qquad B_{\mu\nu}=g_{\mu\nu}+2 \ap
k_{\mu}k_{\nu}\qquad,\qquad H_{\mu}=2\ap k_{\mu}\, .\end{equation}
As a result, we obtain one complex scalar $\Phi^a\equiv\Phi^{a+}$
($\ov\Phi^a\equiv\Phi^{a-})$
and one $J=2$ particle $B^a$, with the vertices given  by
\begin{eqnarray}\nonumber
V_{\Phi^{a\pm}}^{(-1)}(z,k) &=& \frac{g_A}{2\sqrt{2\alpha'}}\, T^a\, e^{-\phi}
\,
\Bigl[ \, (g_{\mu\nu}
\, + \, 2\ap k_{\mu}k_{\nu})\, i\partial X^{\mu} \, \psi^{\nu} \ + \ 2\ap
k_{\mu}\partial\psi^{\mu} \Bigr. \notag \\
& & \ \ \ \ \ \ \Bigl. \ \pm \  \frac{i}{6}\, 2 \ap \,
\epsilon_{\mu\nu\rho\lambda}\,k^{\lambda} \,\psi^{\mu} \, \psi^{\nu} \,
\psi^{\rho} \, \Bigr] \,e^{ikX}\ ,\label{vertphi}\\[2mm]
V_{B^a}^{(-1)}(z,\alpha,k) &=& \frac{g_A}{\sqrt{2\alpha'}}\, T^a\,
e^{-\phi}\,\alpha_{\mu\nu}\,i\partial
X^{\mu}\psi^{\nu}
\,e^{ikX},\label{vertb}
\end{eqnarray}
The relative normalization of the scalar and pseudoscalar parts of the $J=0$
vertex will become clear later.

On the D--brane world--volume half of the space--time bulk SUSY is realized.
The space--time SUSY currents contain the internal fields $\Si^I,\ov\Si^I$
in the Ramond sector of the internal SCFT with the index $I$ specifying the
SUSY.
In type I or type IIB orientifolds D9--branes wrapped on a CYM
yield $\Nc=1$ world--sheet SUSY. This SUSY gives rise to the pair
$\Si,\ov\Si$ of internal fields whose operator product expansions (OPEs) are
given by \cite{Banks:1987cy}
\bea\label{OPE1}
\Sigma(z)\ \ov \Sigma(w)&=&(z-w)^{-3/4}\ I+(z-w)^{1/4}\
\h\Jc(w)+\ldots\ ,\\[2mm]
\Sigma(z)\ \Sigma(w)&=&(z-w)^{3/4}\ \Oc(w)+\ldots\ ,
\eea
with the dimension one current $\Jc$ and dimension $3/2$ operator $\Oc$.
These fields can be expressed in terms of a canonically normalized free scalar
field $H$
as $\Jc=i\sqrt 3\ \p H$ and $\Oc=e^{i\sqrt 3 H}$, respectively.
For the Ramond field $\Si$, which has charge $3/2$ under the current $\Jc$,
we have $\Si=e^{i\fc{\sqrt3}{2}H}$.
On the other hand, e.g. for D7--branes wrapped on 4--cycles we have $\Nc=2$
space--time
SUSY on their world--volume while for space--time filling D3--branes $\Nc=4$
SUSY.
These extended SUSY algebras give rise to the Ramond fields
$\Si^I,\ov\Si^I$ whose OPEs are given by \cite{Banks:1988yz,Ferrara:1989ud}
\bea\label{OPE2}
\Sigma^I(z)\ \ov \Sigma^J(w)&=&(z-w)^{-3/4}\ \delta^{IJ}\ I+(z-w)^{1/4}\
\Jc^{IJ}(w)+\ldots\ ,\\[2mm]
\Sigma^I(z)\ \Sigma^{J}(w)&=&(z-w)^{-1/4}\ \psi^{IJ}(w)+(z-w)^{3/4}\
\Oc^{IJ}(w)+\ldots\ ,
\eea
with the dimension one currents $\Jc^{IJ}$, the dimension $1/2$ operators
$\psi^{IJ}$
and the dimension $3/2$ operators $\Oc^{IJ}$.

{}From the fact that $\Phi(J=0)$ and $B(J=2)$ resonances propagate in the
universal amplitudes we can conclude that they must remain in the spectra of all
self--consistent compactifications. In the context of SUSY, we can also ask
whether these particles appear in two--gluino annihilations channels and in
general, if there are other massive NS bosons that can be produced as a result
of gluino scattering processes. These bosons are necessary to complete
SUSY multiplets. In the $\Nc=1$ case, only one gluino species, say
$\lambda^1\equiv\lambda$, remains in the spectrum, with a single internal
Ramond
field $\Sigma^1\equiv\Sigma$ in the  vertex operator (\ref{vertlam}). If two
gluinos
are in the same helicity state, the internal charge conservation requires
in the third (NS) vertex operator an internal field of charge $3$ under the
current $\Jc$, therefore they cannot
couple to $\Phi$ or $B$. In this case, they can couple to one complex scalar
field $\Omega$ only, with the vertex operator
\begin{equation}\label{vertom}
V_{\Omega^a}^{(-1)}(z,k) = g_A\ T^a\ e^{-\phi}\ \Oc\ e^{ikX}\ ,
\end{equation}
with the charge $3$ field $\Oc$ appearing in the OPE \req{OPE1}.
The field $\Om$ is  universal to all CY compactifications.
Similarly, for extended space--time SUSY the fields $\Oc^{IJ}$ of the OPE
\req{OPE2} appear in the vertex operator \req{vertom}. In the $\Nc=4$ case
these fields $\Oc^{IJ}$ are essentially represented by
products of three internal fermions  $\Psi^i\Psi^j\Psi^k$.

On the other hand, if
two gauginos carry opposite helicity, then the vertex operator of the third
particle
must contain an odd number (one or three) of worldsheet space--time fermions
$\psi^\mu$ in order to get a non--vanishing correlator, c.f.
\cite{Hartl:2009yf}.
By using explicit forms of the underlying worldsheet correlators, it is easy
to show that the coupling to scalar $\Phi$ vanishes on--shell, but there is a
non--vanishing coupling to $B$. In addition, there is a non-vanishing coupling
to
a vector particle $W(J=1)$, universal to all $\Nc=1$ compactifications, with the
vertex
\begin{equation}\label{vertw}
V_{W^a}^{(-1)}(z,\xi,k) = g\ \sqrt{\frac{\al'}{6}}\ T^a\ e^{-\phi}\
\xi_{\mu}\ \psi^{\mu}\ \Jc\ e^{ikX}\ ,
\end{equation}
with $\cal J$ the dimension one worldsheet current appearing in the OPE
\req{OPE1}.
Similarly, for extended space--time SUSY the currents $\Jc^{IJ}$ of the OPE
\req{OPE2} appear in the vertex operator \req{vertw}.
Thus the universal part of the first massive level of the NS sector in $\Nc=1$
superstring compactifications consists of $B(J=2)$, $W(J=1)$, $\Phi(J=0)$ and
$\Omega(J=0)$.  Together with one massive spin 3/2 (Rarita-Schwinger) fermion
from the R sector, $B(J=2)$ and $W(J=1)$ form one massive spin two
supermultiplet \cite{Berkovits:1997zd}. In addition, two complex scalars
$\Phi(J=0)$ and $\Omega(J=0)$ combine with a spin 1/2 Dirac fermion to form one
massive scalar supermultiplet.

\subsection{\label{fermions}Fermions (R sector)}

\noindent
Also for the fermions, we begin with the first massive level in $D=10$. In the R
sector, the fermion vertex operator [in its canonical $(-1/2)$-ghost picture] is
parametrized by two vectors, Majorana-Weyl spinors $v_m^A$ and $\bar
\rho^n_{\dot B}$ of opposite chirality \cite{Koh:1987hm}:
\begin{equation}
V_{R,a}^{(-1/2)}(z,v, \bar{\rho},k) \ \ = \ \ C_{\Lambda} \  T^a \  \bigl[
\,  v_m^{A} \  i\pa X^m \ + \ 2\al' \, \bar{\rho}^m_{\dot B} \ \psi_m\, \psi^n
\, \Ga_n^{\dot B A} \, \bigr] \,  \Theta_{A}  \  e^{-\phi/2} \ e^{ik  X} \ .
\label{massR}
\end{equation}
Here, $A$ denotes a left-handed spinor index while $\dot B$ is its right handed
counterpart. $\Ga_n$ are $16 \times 16$ Weyl blocks of the $D=10$ gamma matrices
and $\Theta_A$ are the conformal weight $h=\frac{5}{8}$ chiral spin fields.

{}Requiring BRST invariance imposes two on-shell constraints on $v_m^A$ and
$\bar \rho^m_{\dot B}$ which determine $\bar \rho$ in terms of $v$ and leave 144
independent components in the latter. Furthermore, a set of 16 spurious states
exists which allows to take $\bar \rho$ and $v$ as transverse and
$\Ga$-traceless:
\begin{equation}
k^m \, v_m^A \eq v_m^A \, \Ga^m_{ A \dot B} \eq k_m \, \bar \rho^m_{\dot B} \eq
\bar \rho^m_{\dot B} \, \Ga_m^{\dot BA} \eq 0\ .
\end{equation}
These $128=144-16$ physical degrees of freedom match the counting for bosons.

{}In $D=4$ compactifications, the spin field $\Theta$ breaks into the products
of $SO(1,3)$ covariant spin fields $S_\al, \bar S^{\dbe}$ times the internal
spin fields of weight $h=\frac{3}{8}$. Our goal is to construct the vertex for
the color triplets $Q$, the first excited level of quarks originating from brane
intersections
[created by vertex operators written in Eq. (\ref{vertq})]. To
that end, we start
from the $D=4$ decomposition of the vertex (\ref{massR}) and combine
$h=\frac{5}{4}$ conformal fields of the spacetime SCFT (namely $\pa X^\mu S_\al$
and $\psi^\mu \psi^\nu S_\al$) with the $h=\frac{3}{8}$ fermionic boundary
changing operator $\Xi^{a \cap b}$ which plays the role  of  internal spin
field.
Finally, we multiply by the $h=\frac{3}{8}$ superghost and the $h= -1$
exponential $e^{ikX}$ factors to obtain the  desired $h=1$. In this way, we
obtain:
\bea\label{massq}
V_{Q^{\al}_{\beta}}^{(-1/2)}(z,v, \bar{\rho},k) &=&{\al'}^{1/4} e^{\phi_{10}/2}
\
\bigl(T^{\al}_{\beta} \bigr)^{\beta_{1}}_{\al_{1}} \  \bigl[ \, i v_\mu^{\beta}
\, \pa X^\mu  -\sqrt{\al'} \, \bar{\rho}^\mu_{\dal} \, \psi_\mu \, \psi^\nu \,
\bar
\si_\nu^{\dal \beta} \, \bigr] \,  S_{\beta}\    e^{-\phi/2}\ \Xi^{a \cap b} \
e^{ik  X}\ , \\[1mm]
V_{\bar{Q}^{\beta}_{\al}}^{(-1/2)}(z,\bar{v}, \rho ,k) &=&{\al'}^{1/4}
e^{\phi_{10}/2}  \
\bigl(T^{\beta}_{\al} \bigr)^{\al_{1}}_{\beta_{1}} \ \bigl[ \, i
\bar{v}^\mu_{\dbe} \, \pa X_\mu  -\sqrt{\al'} \, \rho_\mu^\al \, \psi^\mu \,
\psi_\nu \, \si^\nu_{\al \dbe} \, \bigr] \,  S^{\dbe}\   e^{-\phi/2}\
\bar{\Xi}^{a
\cap b} \  e^{ik X }\ .
\eea

The on-shell BRST constraints can be derived by following the analysis of
Ref. \cite{Koh:1987hm}, although some numerical coefficients need adjustment
from
$D=10$ to $D=4$, due to gamma matrix contractions such as $\Ga^{\mu \nu} \Ga_\nu
= -(D-1) \Ga^\mu$. They are:
\begin{equation}
k_\mu \, \bar v^\mu_{\dal} \ \ = \ \ -\frac{3}{4\sqrt{\al'}} \; \rho_\mu^\beta
\,
\si^\mu_{\beta \dal} \co \rho^{ \al}_{\mu} \ \ = \ \sqrt{\al'} \; \bar
v_{\mu \dbe} \, \bar \si_{\nu}^{ \dbe \al} \, k^\nu \ + \ \frac{1}{2} \;
\rho_\nu^\beta \, (\si^\nu \, \bar \si_\mu )_\beta \, \! ^{\al}\ .
\label{onshell2}
\end{equation}
The above constraints allow expressing $\rho$ in terms of $\bar v$
\begin{equation}
\rho^{\mu \al} \ = \sqrt{\al'} \;  \bar v^{\mu}_{ \dbe} \, \bar \si_{\nu}^{
\dbe \al} \, k^\nu \ - \ \frac{2}{3}\sqrt{\al'} \, k_\nu \, \bar v^\nu_{\dbe} \,
\bar
\si^{\mu \dbe \al}
\label{onshell6},
\end{equation}
with $\bar v$ subject to:
\begin{equation}
 \bar v_{\dal}^\mu \,  \bar \si_\mu^{\dal \beta}  = 2 \alpha' \, k_\mu \, \bar
v^\mu_{\dal} \,
 \bar\si^{\dal \beta} _\nu \, k^\nu \ .
\label{onshell7}
\end{equation}
The latter impose 2 constraints on 8 (complex) parameters, leaving 6 wave
functions describing spin 3/2
and spin 1/2 resonances $Q^{\star}(J=3/2)$ and $Q(J=1/2)$, respectively.

Next, we disentangle the different spin components of the solutions of
Eq. (\ref{onshell7}). A general vector spinor of $SO(1,3)$ can be decomposed
into
irreducible representations of spin $3/2$ and $1/2$ according to $\bigl( \bf
\frac{1}{2} , \bf \frac{1}{2} \bigl) \otimes \bigl( \bf 0, \bf \frac{1}{2}
\bigl) = \bigl(  \bf \frac{1}{2}, \bf 1 \bigl) \oplus \bigl( \bf \frac{1}{2} ,
\bf 0 \bigl)$. The spin $3/2$ states $\bar \chi \in  \bigl( \bf 1 , \bf
\frac{1}{2} \bigl)$ can be extracted by imposing the irreducibility conditions
\begin{equation}
\te{spin} \ \tfrac{3}{2}  : \ \ \ \bar v^\mu_{\dal}(J=3/2) \ = \  \bar
\chi^\mu_{\dal} \ \ \ \te{such that} \ \ \ \bar \chi^\mu_{\dal} \, \bar
\si_\mu^{\dal \beta} \eq k_\mu \, \bar \chi^\mu_{\dal} \eq 0\ ,
\label{spin32}
\end{equation}
which do indeed assemble 4 solutions of Eq. (\ref{onshell7}).
In this case, the general formula (\ref{onshell6}) for $\rho$ simplifies since
irreducibility kills the second term on the right hand side:
\begin{equation} \te{spin} \ \tfrac{3}{2}  : \ \ \
\rho^{\mu \al}(J=3/2)\ = \eta^{\mu \al} \eq \sqrt{\al'} \;  \bar
\chi^{\mu}_{ \dbe} \, \bar \si_{\nu}^{ \dbe \al} \, k^\nu\ .
\label{Dirac}
\end{equation}
The latter can be recast as a massive Dirac equation for the four-spinor built
from $\eta\oplus \bar \chi$:
\begin{equation}
\ccb 0 &\not \! k_{\al \dbe} \ \\ \not \! k^{\dal \beta} &0 \cce \, \vecb
\eta_{\mu\beta} \\  \, \bar \chi_\mu^{\dbe}\vece \eq - \, m
\vecb  \eta_{\mu\alpha} \\  \, \bar \chi_\mu^{\dal}\vece \co
m \eq \frac{1}{\sqrt{\al'}}\ .
\label{Dirac2}
\end{equation}

The remaining 2 solutions of Eq. (\ref{onshell7}) describe a spin 1/2 particle
and  can be parametrized in terms of a left-handed  Weyl spinor $\eta\in \bigl(
\bf
\frac{1}{2} , \bf 0 \bigl)$:
\begin{equation}\bar v^\mu_{\dbe}=
\eta^\al \si^\nu_{\al \dbe} \bigl[ \de^\mu_\nu +
2 \alpha' k^\mu k_\nu \bigr]\ .
\end{equation}
Incidentally, in $D=10$, similar solutions of the BRST conditions turned out to
be spurious, but in any lower dimensions, they belong to the physical spectrum.
This ties in with the observations made for real part of the massive scalar
(\ref{vertb}).
Actually, the spin $1/2$ wave functions can be succinctly parametrized in terms
of the right handed counterpart  $\bar \chi_{\dbe}$ associated to $\eta$ via
massive
Dirac equation, $\bar \chi_{\dbe} = \sqrt{\ap} \, \eta^\al \!\not \! k_{\al
\dbe}$:
\begin{equation}
\te{spin} \ \tfrac{1}{2}  : \ \ \ \bar v^\mu_{\dal}(J=1/2) \eq - \,
\frac{\sqrt{\al'}}{2\sqrt{2}}
\; \bar\chi_{\dbe} \, (\bar \si^{\mu } \! \not \! k)^{\dbe} \, \! _{ \dal} \co
\rho_\mu^{\al}(J=1/2) \eq -\frac{1}{6\sqrt{2}} \; \bar\chi_{\dbe} \, \bar
\si_\mu^{\dbe \al}.
\label{spin12}
\end{equation}

{} Finally, we should mention that the vertex operators of excited gauginos can
be obtained from the massive quarks (\ref{massq})  by
adjusting normalization, Chan-Paton factors and the internal spin field:
\begin{equation}
\te{quarks:} \ \bigl( C_Q , T^\al_\beta , \Xi^{a \cap b} \bigr) \ \to \
\te{gluinos:} \ \bigl( C_\Lambda, T^a ,\Si^I \bigr) \ .
\label{gluino}
\end{equation}
These massive quarks (and gluinos) couple to their massless progenitors and
gluons. In particular, $Q^{\star}(J=3/2)$ and $Q(J=1/2)$ propagate in the
gluon-quark
fusion channels of universal four-point amplitudes. Both spins present in the
composite vertex for massive fermions belong to the spectrum regardless of the
compactification geometry.

\section{Weyl--van--der--Waerden formalism for wave functions of massless and massive particles}
The vertex operators described in the previous Section depend on the wave
functions of created string states.  {}For our purposes, it is very
convenient to use a unified ``helicity notation'', \ie the Weyl--van--der--Waerden
formalism for both massless and massive particles. To that end, we use the
notation of Wess and Bagger \cite{Wess:1992cp}, with the metric signature
$(-,+,+,+)$, however our conventions are slightly different than in the
Refs. \cite{Lust:2008qc,Lust:2009pz},
therefore we phase them in gradually, in the context of wave functions
describing massless particles.

\subsection{Massless spin 1/2 and 1  Wave Functions}
\noindent
Massless spin 1/2 fermions are usually described by two-component Weyl
spinors, however anticipating the massive case, it is convenient to recall
the four-component $\bigl( \bf
\frac{1}{2} , \bf \frac{1}{2} \bigl)$ representation of the wave functions
$u_\pm(k)$ describing  their $\pm\frac{1}{2}$ helicity eigenstates,
respectively:
\begin{eqnarray}u_{+}(k)
&=&
|k\rangle=\binom{0}{k^{*\dot{a}}}\ ,\nonumber
\end{eqnarray}
\begin{eqnarray}
u_{-}(k)&=&
 |k]\,=\,\binom{k_{a}}{\,0\,  }\, .\label{umass}
\end{eqnarray}
The momentum  $k$ is on-shell, $k^2=0$, and can
be factorized as:
\begin{equation}
k_{\mu}\sigma_{a\dot{a}}^{\mu}=k_{a\dot{a}}=-k_{a}k_{\dot{a}} ^*\ .
\end{equation}
Note that we are using
$a$ and $\dot{a}$, as well as other letters from the beginning
of the Latin alphabet, as $SU(2)\times SU(2)$ spinor indices that can be raised
or lowered in the standard way by the $\epsilon$-symbol. The conjugate spinors
are:
\begin{eqnarray}\bar{u}_{+}(k)
&=&
[k|=\,\left(k^{a},\,0\right)\ ,\nonumber \\[1mm]
\bar{u}_{-}(k)&=&
 \langle k|=\,\left(k_{\dot{a}}^*,0\right)\ .
\end{eqnarray}
Our conventions for the spinor products follow Ref. \cite{Dixon:1996wi}:
\begin{eqnarray}
\langle pq\rangle &=&
\langle p|q\rangle=\bar{u}_{-}(p)u_{+}(q)~=~p_{\dot{a}}^{*}q^{*\dot{a}}\ ,
\nonumber\\
{}[pq]&=&[p|q]\, =\bar{u}_{+}(p)u_{-}(q)~=~p^{a}q_{a}\ ,\end{eqnarray}
so that:
\begin{equation}
\langle pq\ \rangle[qp]=-2pq\ .\end{equation}

The polarization vectors for a massless spin 1 particle with momentum $k$
utilize an arbitrary gauge-fixing ``reference''momentum $r$ for each gauge
boson \cite{Dixon:1996wi}, which can be chosen to be any light-like momentum
except $k$. They can be written as:
\begin{eqnarray}
\epsilon_{\mu}^{+}(k,r)&=&~\frac{\langle
r|\gamma_{\mu}|k]}{\sqrt{2}\langle
rk\rangle}~=~ \frac{r_{\dot{a}}^{*}\bar{\sigma}_{\mu}^{\dot{a}a}k_{a}}{\sqrt{2}
\langle rk\rangle}\ ,\nonumber\\
\epsilon_{\mu}^{-}(k,r)&=&-\frac{[r|\gamma_{\mu}|k\rangle}{
\sqrt{2}[rk]}=-\frac{r^{a}\sigma_{\mu
a\dot{a}}k^{*\dot{a}}}{\sqrt{2}[rk]}\label{eps0}\ .\end{eqnarray}
Of course, the physical amplitudes must not depend on the choice of reference
momenta.

\subsection{Massive spin 1 and spin 2 bosons}
\noindent
A spin $J$ particle contains $2J+1$ spin degrees of freedom associated to the
eigenstates of $J_z$. The choice of the quantization axis $z$ can be handled in
an elegant way by decomposing the momentum $k$ into {\it two} arbitrary
light-like reference momenta $p$ and $q$:
\begin{equation} k^{\mu}=p^\mu+q^{\mu}\ ,\qquad k^2=-m^2=2pq\
,\qquad p^2=q^2=0\ .\end{equation}
Then the spin quantization axis is chosen as the direction of $p$ in the
rest frame. The $2J+1$ spin wave functions depend of $p$ and $q$
\cite{Novaes:1991ft, Spehler:1991yw}, however this dependence drops out in the
amplitudes
summed over all spin directions  and in ``unpolarized'' cross sections.

The massive spin 1 wave functions $\xi^\mu$ are given by the following
(transverse, \ie $k_\mu\xi^\mu=0$) polarization vectors:
\begin{eqnarray}
\xi^{\mu}(k, +1) &
=&\frac{\sqrt{2}}{m}p_{\dot{a}}^{*}\bar{\sigma}^{\mu\dot{a}a}q_{a}\ ,\nonumber\\
\xi^{\mu}(k, ~0\,)~ &
=&\frac{1}{m}\bar{\sigma}^{\mu\dot{a}a}(p_{\dot{a}}^{*}p_{a}-q_{\dot{a}}^{*}q_{a
}
)\label{xi1}\ ,\\
\xi^\mu(k, -1) &
=&\frac{\sqrt{2}}{m}q_{\dot{a}}^{*}\bar{\sigma}^{\mu\dot{a}a}p_{a}\ .\nonumber
\end{eqnarray}

The massive spin 2 wave functions are the traceless, symmetric tensors
$\alpha^{\mu\nu}$ ($\alpha^{\mu\nu}=\alpha^{\nu\mu},
~g_{\mu\nu}\alpha^{\mu\nu}=0$), subject to the transversality constraint
$k_\mu\alpha^{\mu\nu}=0$. They can be written as \cite{Spehler:1991yw}:
\begin{eqnarray}
{\alpha}^{\mu\nu}(k,+2) &
=&\frac{1}{2m^{2}}\bar{\sigma}^{\mu\dot{a}a}\bar{\sigma}^{\nu\dot{b}b}p_{\dot{a}
}^{*}q_{a}p_{\dot{b}}^{*}q_{b}\ ,\nonumber\\
\alpha^{\mu\nu}(k,+1) &
=&\frac{1}{4m^{2}}\bar{\sigma}^{\mu\dot{a}a}\bar{\sigma}^{\nu\dot{b}b}\left[(p_{
\dot{a}}^{*}p_{a}-q_{\dot{a}}^{*}q_{a})p_{\dot{b}}^{*}q_{b}+p_{\dot{a}}^{*}q_{a}
(p_{\dot{b}}^{*}p_{b}-q_{\dot{b}}^{*}q_{b})\right]\ ,\nonumber\\
\alpha^{\mu\nu}(k,~0~) &
=&\frac{1}{2m^{2}\sqrt{6}}\bar{\sigma}^{\mu\dot{a}a}\bar{\sigma}^{\nu\dot{b}b}
\left[(p_{\dot{a}}^{*}p_{a}-q_{\dot{a}}^{*}q_{a})(p_{\dot{b}}^{*}p_{b}-q_{\dot{b
}}^{*}q_{b})-p_{\dot{a}}^{*}q_{a}q_{\dot{b}}^{*}p_{b}-q_{\dot{a}}^{*}p_{a}p_{
\dot{b}}^{*}q_{b})\right]\ ,\nonumber\\
\alpha^{\mu\nu}(k,-1) &
=&\frac{1}{4m^{2}}\bar{\sigma}^{\mu\dot{a}a}\bar{\sigma}^{\nu\dot{b}b}\left[(q_{
\dot{a}}^{*}q_{a}-p_{\dot{a}}^{*}p_{a})q_{\dot{b}}^{*}p_{b}+q_{\dot{a}}^{*}p_{a}
(q_{\dot{a}}^{*}q_{a}-p_{\dot{b}}^{*}p_{b})\right]\ ,\label{al1}\\
\alpha^{\mu\nu}(k,-2) &
=&\frac{1}{2m^{2}}\bar{\sigma}^{\mu\dot{a}a}\bar{\sigma}^{\nu\dot{b}b}q_{\dot{a}
}
^{*}p_{a}q_{\dot{b}}^{*}p_{b}\ .\nonumber\end{eqnarray}

\subsection{Massive spin 1/2 and 3/2 fermions}
\noindent
Massive spin 1/2 wave functions are represented by four-component spinors
\begin{equation}
U(k)=\binom{\eta_{a}}{\bar{\chi}^{\dot{a}}}\ ,\label{u0}\end{equation}
with the upper and lower components related by the Dirac equation
\begin{equation}
k_{\mu}\bar{\sigma}^{\mu\dot{a}a}\eta_{a}=-m\bar{\chi}^{\dot{a}}\qquad
({\it i.e}.\\ k_{\mu}\sigma_{a\dot{a}}^{\mu}\bar{\chi}^{\dot{a}}=-m\eta_{a})\
.\label{dirdir}\end{equation}
Their explicit form is:
\begin{eqnarray}
U(k,+\frac{1}{2}) & =&\binom{\frac{\langle
qp\rangle}{m}q_{a}}{p^{*\dot{a}}}\ ,\nonumber\\[3mm]
U(k, -\frac{1}{2}) &
=&\binom{p_{a}}{\frac{[qp]}{m}q^{*\dot{a}}}\label{u1}\ .\end{eqnarray}
These wave functions should be substituted to Eqs. (\ref{spin12})
and (\ref{massq}) in order to obtain the vertex for the $Q(J{=}1/2)$
resonance. Actually, Eq. (\ref{spin12}) utilizes only the lower component
\nolinebreak $\bar\chi^{\dot{a}}$.

The wave functions of massive spin 3/2 particles are the vector spinors
\begin{equation}
R^\mu(k)=\binom{\eta_{a}^\mu}{\bar{\chi}^{\mu\dot{a}}}\
,\label{umu1}\end{equation}
with the vectors constrained by
\begin{equation}
k_\mu\chi^{\mu\dot{a}}=k_\mu\eta^{\mu}_a
=\sigma_{a\dot{a}}^{\mu}\bar{\chi}^{\mu\dot{a}}=\bar{\sigma}^{\mu\dot{a}a}\eta_{
a}^\mu=0
\end{equation}
and satisfying the Dirac equation (\ref{dirdir}). The wave functions describing
individual spin configurations are
\cite{Novaes:1991ft}:
\begin{eqnarray}
R^{\mu}(k,+\frac{3}{2}) & =&\frac{1}{\sqrt{2}m}\binom{\frac{\langle
qp\rangle}{m}q_{a}}{p^{*\dot{a}}}(p_{\dot{b}}^{*}\bar{\sigma}^{\mu\dot{b}b}q_{b}
)\ ,\nonumber\\
R^{\mu}(k,+\frac{1}{2}) &
=&\frac{\bar{\sigma}^{\mu\dot{b}b}}{\sqrt{6}m}\left[\binom{\frac{\langle
qp\rangle}{m}q_{a}}{p^{*\dot{a}}}(p_{\dot{b}}^{*}p_{b}-q_{\dot{b}}^{*}q_{b}
)+\binom{\frac{\langle
qp\rangle}{m}p_{a}}{-q^{*\dot{a}}}(p_{\dot{b}}^{*}q_{b})\right]\ ,\nonumber\\
R^{\mu}(k,-\frac{1}{2}) &
=&\frac{\bar{\sigma}^{\mu\dot{b}b}}{\sqrt{6}m}\left[\binom{p_{a}}{\frac{[qp]}{m}
q^{*\dot{a}}}(p_{\dot{b}}^{*}p_{b}-q_{\dot{b}}^{*}q_{b})+\binom{-q_{a}}{\frac{[
qp]}{m}p^{*\dot{a}}}(q_{\dot{b}}^{*}p_{b})\right]\ ,\label{umu2}\\
R^{\mu}(k,-\frac{3}{2}) &
=&\frac{1}{\sqrt{2}m}\binom{p_{a}}{\frac{[qp]}{m}q^{*\dot{a}}}(q_{\dot{b}}^{*}
\bar{\sigma}^{\mu\dot{b}b}p_{b})\ .\nonumber\end{eqnarray}
These wave functions should be substituted to Eqs. (\ref{spin32}), (\ref{Dirac})
and (\ref{massq}) in order to obtain the vertex for the $Q^{\star}(J=3/2)$
resonance. Here again, only the lower component
$\bar\chi^{\mu\dot{a}}$ enters into the vertex explicitly.

\section{Two-- and three--particle decay amplitudes}

With the vertices and wave functions at hand, we
are ready to compute the amplitudes\footnote{For a recent review on
three-- and four--point amplitudes involving higher spin states in bosonic
uncompactified string theory, see Ref. \cite{Sagnotti:2010at}.} describing two--
and three--particle decays
of the bosons, $B(J=2)$, $W(J=1)$, $\Phi(J=0)$, $\Omega(J=0)$
and fermions $Q(J=1/2)$, $Q^{\star}(J=3/2)$. Although we are mainly
interested in color octet bosons and fermion triplets, we will be considering
bosons in the adjoint representation of a general $U(N)$ gauge group and
fermions in its fundamental $N$ representation. Technical
aspects of such computations have been discussed at length in
Refs. \cite{Lust:2008qc} and \cite{Lust:2009pz},  therefore we
simply state the results, and go into more details only if the computations are
complicated or they involve some new elements. Some amplitudes
involving massive string states have been discussed before in
Ref. \cite{Liu:1987tb}.

Two-particle decays described by the amplitudes involving one massive state and
two massless particles are particularly simple because the underlying SCFT
correlators of three vertex operators inserted at the disk boundary do not
depend on the vertex positions, therefore integrals over these positions are
trivial. They are important, however, for the determination of proper
normalization  of vertex operators, which is done by
factorizing a four-point scattering amplitude of massless particles at the
resonance pole and comparing with appropriate products of three-point
amplitudes. The vertex operators written in Section III have been already
normalized in this way, so the amplitudes written below can serve as a check,
as illustrated below in some more interesting cases.

We will be using the following shorthand notation for the amplitudes
$$R[J_z;\lambda_1, \lambda_2, \dots]$$ denotes the amplitude for the decay of
Regge state $R(=B,W,\dots)$  with the spin $z$-component $J_z$ into massless
particles (quarks or gluons) labeled by $i=1,2,\dots$, with the momenta $k_i$,
helicities $\lambda_i$ (they can be also specified by spinors or polarization
vectors), {\it etc}.
The massive particle will be labeled by the
last index, \ie $i=3$ in three-point functions and $i=4$ in
four-point functions.
The Mandelstam variables are defined as
\begin{equation}\label{mandel}
 s=-(k_1+k_2)^2\ ,\qquad  t=-(k_1+k_3)^2\ ,\qquad u=-(k_1+k_4)^2\end{equation}
with all  momenta incoming and on-shell:
\begin{equation}\sum_{i=1}^{4}k_i=0\ ,\qquad
k_1^2=k_2^2=k_3^2=0\ ,\qquad k^2_4=-m^2=-\textstyle\frac{1}{\alpha'}\ ,
\end{equation}
which implies the following relation for the dimensionless variables:
\begin{equation}
 \alpha's+\alpha't+\alpha'u=1\label{stu}\ .
\end{equation}
The spinor products will be abbreviated as
\begin{equation}
\langle k_i|k_j\rangle =\langle ij\rangle\ ,\qquad [ k_i|k_j]
=[ ij]\ .
\end{equation}
Finally, we recall the string formfactor
\begin{equation}
 V_t=V(s,t,u)=\frac{\Gamma(1-\alpha's)\Gamma(1-\alpha'u)}{
\Gamma(1-\alpha's-\alpha'u)}\ ,\qquad V_s=V_t(t\leftrightarrow s)\
,\quad V_u=V_t(t \leftrightarrow u)\ .
\end{equation}
Note that once the kinematic constraint (\ref{stu}) is implemented,
\begin{equation}
 V_t=\frac{\Gamma(1-\alpha's)\Gamma(1-\alpha'u)}{
\Gamma(\alpha't)}\ ,
\end{equation}
with the denominator $\Gamma(\alpha't)$ different from $\Gamma(1+\alpha't)$
that appears in the corresponding function describing the universal disk
amplitudes with four massless particles  \cite{Anchordoqui:2008hi}.
CFT correlators involving various NS fermions and R spin fields are determined
using
the results of \cite{Hartl:2009yf}.

\subsection{Massive spin 2 boson $B(J=2)$}

\noindent
We begin with the $B$-decays into gluons. The two-gluon channel is described by
the amplitude
\begin{equation}
B[\alpha;\epsilon_1,\epsilon_2]=(2\,d^{a_1a_2a_3})\ (4g\sqrt{2\al'})\
\alpha_ { \mu\nu }
\left [ \, (\epsilon_2 k_{1})k_{2}^{\mu}\epsilon_{1}^{\nu}+(\epsilon_{1}
k_{2})k_{1}^{\mu}\epsilon_{2}^{\nu}-(k_{1}
k_{2})\epsilon_{2}^{\mu}\epsilon_{1}^{\nu}-(\epsilon_{1}\epsilon_{2})k_{1}^
{\mu}k_{2}^{\nu}\,\right]\ .\label{bamp1}\end{equation}
In the prefactor, we singled out the color factor,
\begin{equation}
 2\,d^{a_1a_2a_3}={\rm Tr}(T^{a_1}T^{a_2}T^{a_3})+{\rm
Tr}(T^{a_2}T^{a_1}T^{a_3})\ ,\label{colfac}
\end{equation}
which appears after adding the contributions of the two orderings of the vertex
operators inserted at the disk boundary. Note, that the form of the
vertex \req{bamp1} is reminiscent to the generic vertex \req{Massive2}.
It is convenient to rewrite the
amplitude (\ref{bamp1}) as
\begin{equation}
 B[\alpha;\epsilon_1,\epsilon_2]=4g\
 (2\,d^{a_1a_2a_3})\ (2\al')^{3/2}\ \mathscr{B} \left[
\alpha;\epsilon_1,\epsilon_2\right]\label{bamp2}
\end{equation}

In order to rewrite the above amplitude in the helicity basis, for gluon
helicity eigenstates, we substitute the wave functions (\ref{eps0}) and
(\ref{al1}). We find
\begin{equation}
\mathscr{B}[\alpha;\pm,\pm]=0\ ,
\end{equation}
thus non-vanishing amplitudes must necessarily involve two gluons with
opposite polarizations. They read:
\begin{eqnarray}
\mathscr{B}\left[-2;+,-\right] &
=&-\frac{1}{4}\langle p2\rangle^2[q1]^{2}\ ,\nonumber\\
\mathscr{B}\left[-1;+,-\right] &
=&-\frac{1}{2}\langle p2\rangle^{2}[q1][p1]\ ,\nonumber\\
\mathscr{B}\left[~0~;+,-\right] & =&-\frac{\sqrt{6}}{4}
\langle p2\rangle^2[p1]^{2}\ ,\label{bamp3}\\
\mathscr{B}\left[+1;+,-\right] & =
&+\frac{1}{2}\langle p2\rangle\langle q2\rangle[p1]^{2}\ ,\nonumber\\
\mathscr{B}\left[+2;+,-\right] & =
&-\frac{1}{4}\langle q2\rangle^2[p1]^{2}\ .\nonumber
\end{eqnarray}

As a first check of the above result, we can examine the probability for the
decay of unpolarized $B$ into
a specific $(+,-)$ helicity configuration, by computing the sum
\begin{equation}
 \sum_{\alpha=-2}^{+2}|B(\alpha;+,-)|^2=\frac{8}{\al'}\, g^2
(2\,d^{a_1a_2a_3})^2\ ,\label{prob2g}
\end{equation}
which does indeed turn out to be independent of the choice of reference
vectors $p$ and $q$. Now we can check if the result is consistent with string
factorization. {}From Ref. \cite{Anchordoqui:2008hi} we know that only the spin
2
resonance
appears in the $s$-channel of the four-gluon amplitude ${\cal
M}[g_1^+,g_2^-,g_3^-,g_4^+]$, where it yields the following
residue at $s=M^2=1/\al'$
\begin{equation}
{\rm Res}_{s=1/\ap}\ {\cal M}[g_1^+,g_2^-,g_3^-,g_4^+] \eq  4\,g^2 {\rm
Tr}
(T^{a_1}T^{a_2}T^{a_3}T^{a_4})\,\al'\langle 23\rangle^2 [14]^2
+\dots\ , \label{mfact}
\end{equation}
where we picked up just one partial amplitude contribution. In order to compare
our $B$-decay amplitude with the residue, we compute
\begin{equation}
  \sum_{\alpha=-2}^{+2}B(\alpha;+,-)\ [B(\alpha;+,-)|_{(1\to 3,2\to 4)}]^*\
,\label{sumb}
\end{equation}
with the color factor associated to the first ordering in Eq. (\ref{colfac}),
c.f. Eq. \req{completeness}.
The simplest way to perform the sum (\ref{sumb}) is to set $p=k_1$ and $q=k_2$
because then only $J_z=-2$ contributes. Indeed, after combining the spin and
color sums we recover Eq. (\ref{mfact}), thus confirming the correct
normalization of the vertex operator (\ref{vertb}). Eqs. (\ref{bamp2}) and
(\ref{bamp3}) can be also checked by comparing directly with Eq. (25) of
Ref. \cite{Anchordoqui:2008hi}.

Three-gluon $B$--decays are described by the following amplitude
\begin{align}
B&[\alpha;\epsilon_1,\epsilon_2,\epsilon_3] ~=~ \nonumber
4\, g^2\sqrt{2\al'}\ \big( \, V_t \, t^{a_1a_2a_3a_4} \ + \ V_s \,
t^{a_2a_3a_1a_4}
\ + \ V_u \, t^{a_3a_1a_2a_4} \, \big)\\
&\times \, \biggl\{ \, \frac{1}{s} \; \Bigl[ \, (\epsilon_2 \, \epsilon_3) \,
(\epsilon_1 \,
k_2) \, (k_3^\mu \, \al_{\mu \nu} \, k_3^\nu) \ - \ (\epsilon_1 \, \epsilon_3)
\, (\epsilon_2
\, k_1) \, (k_3^\mu \, \al_{\mu \nu} \, k_3^\nu) \ + \ (\epsilon_1 \,
\epsilon_2) \,
(\epsilon_3 \, k_2) \, (k_1^\mu \, \al_{\mu \nu} \, k_3^\nu)  \Bigr. \biggr.
\notag
\\
& \ \ \ \ \ \ \ \Bigl. - \ (\epsilon_1 \, \epsilon_2) \, (\epsilon_3 \, k_1) \,
(k_2^\mu \,
\al_{\mu \nu} \, k_3^\nu) \ + \ (\epsilon_1 \, k_3) \, (\epsilon_2 \, k_1) \,
(k_3^\mu \,
\al_{\mu \nu} \, \epsilon_3^\nu) \ - \ (\epsilon_2 \, k_3) \, (\epsilon_1 \,
k_2) \, (k_3^\mu
\, \al_{\mu \nu} \, \epsilon_3^\nu) \Bigr. \notag \\
& \ \ \ \ \ \ \ \Bigl. + \ (\epsilon_2 \, k_1) \, (\epsilon_3 \, k_4) \,
(k_3^\mu \,
\al_{\mu \nu} \, \epsilon^\nu_1) \ - \ (\epsilon_1 \, k_2) \, (\epsilon_3 \,
k_4) \, (k_3^\mu
\, \al_{\mu \nu} \, \epsilon^\nu_2) \Bigr. \notag \\
& \ \ \ \ \ \ \ \Bigl. + \ \frac{1}{2\al'} \; (\epsilon_1 \, k_2) \,
(\epsilon^\mu_2 \,
\al_{\mu \nu} \, \epsilon^\nu_3) \ - \ \frac{1}{2\al'} \; (\epsilon_2 \, k_1) \,
(\epsilon^\mu_1 \, \al_{\mu \nu} \, \epsilon^\nu_3) \Bigr. \notag \\
& \ \ \ \ \ \ \ \Bigl. - \ \frac{t}{2} \; (\epsilon_1 \, \epsilon_2) \, (k_2^\mu
\,
\al_{\mu \nu} \, \epsilon_3^\nu) \ + \ \frac{u}{2} \; (\epsilon_1 \, \epsilon_2)
\,
(k_1^\mu \, \al_{\mu \nu} \, \epsilon_3^\nu) \, \Bigr] \notag \\
& \ \ \ + \ \frac{1}{u} \; \Bigl[ \, (\epsilon_1 \, \epsilon_3) \, (\epsilon_2
\, k_3) \,
(k_1^\mu \, \al_{\mu \nu} \, k_1^\nu) \ - \ (\epsilon_1 \, \epsilon_2) \,
(\epsilon_3 \, k_2)
\, (k_1^\mu \, \al_{\mu \nu} \, k_1^\nu) \ + \ (\epsilon_2 \, \epsilon_3) \,
(\epsilon_1 \,
k_3) \, (k_1^\mu \, \al_{\mu \nu} \, k_2^\nu)  \Bigr. \biggr. \notag \\
& \ \ \ \ \ \ \ \Bigl. - \ (\epsilon_2 \, \epsilon_3) \, (\epsilon_1 \, k_2) \,
(k_1^\mu \,
\al_{\mu \nu} \, k_3^\nu) \ + \ (\epsilon_2 \, k_1) \, (\epsilon_3 \, k_2) \,
(k_1^\mu \,
\al_{\mu \nu} \, \epsilon_1^\nu) \ - \ (\epsilon_3 \, k_1) \, (\epsilon_2 \,
k_3) \, (k_1^\mu
\, \al_{\mu \nu} \, \epsilon_1^\nu) \Bigr. \notag
\end{align}

\begin{align}
& \ \ \ \ \ \ \ \Bigl. + \ (\epsilon_3 \, k_2) \, (\epsilon_1 \, k_4) \,
(k_1^\mu \,
\al_{\mu \nu} \, \epsilon^\nu_2) \ - \ (\epsilon_2 \, k_3) \, (\epsilon_1 \,
k_4) \, (k_1^\mu
\, \al_{\mu \nu} \, \epsilon^\nu_3) \Bigr. \notag \\
& \ \ \ \ \ \ \ \Bigl. + \ \frac{1}{2\al'} \; (\epsilon_2 \, k_3) \,
(\epsilon^\mu_1 \,
\al_{\mu \nu} \, \epsilon^\nu_3) \ - \ \frac{1}{2\al'} \; (\epsilon_3 \, k_2) \,
(\epsilon^\mu_1 \, \al_{\mu \nu} \, \epsilon^\nu_2) \Bigr. \notag \\
& \ \ \ \ \ \ \ \Bigl. - \ \frac{s}{2} \; (\epsilon_2 \, \epsilon_3) \, (k_3^\mu
\,
\al_{\mu \nu} \, \epsilon_1^\nu) \ + \ \frac{t}{2} \; (\epsilon_2 \, \epsilon_3)
\,
(k_2^\mu \, \al_{\mu \nu} \, \epsilon_1^\nu) \, \Bigr] \notag \\
& \ \ \ + \ \frac{1}{t} \; \Bigl[ \, (\epsilon_1 \, \epsilon_2) \, (\epsilon_3
\, k_1) \,
(k_2^\mu \, \al_{\mu \nu} \, k_2^\nu) \ - \ (\epsilon_2 \, \epsilon_3) \,
(\epsilon_1 \, k_3)
\, (k_2^\mu \, \al_{\mu \nu} \, k_2^\nu) \ + \ (\epsilon_1 \, \epsilon_3) \,
(\epsilon_2 \,
k_1) \, (k_2^\mu \, \al_{\mu \nu} \, k_3^\nu)  \Bigr. \biggr. \notag \\
& \ \ \ \ \ \ \ \Bigl. - \ (\epsilon_1 \, \epsilon_3) \, (\epsilon_2 \, k_3) \,
(k_1^\mu \,
\al_{\mu \nu} \, k_2^\nu) \ + \ (\epsilon_3 \, k_2) \, (\epsilon_1 \, k_3) \,
(k_2^\mu \,
\al_{\mu \nu} \, \epsilon_2^\nu) \ - \ (\epsilon_1 \, k_2) \, (\epsilon_3 \,
k_1) \, (k_2^\mu
\, \al_{\mu \nu} \, \epsilon_2^\nu) \Bigr. \notag \nonumber\\
& \ \ \ \ \ \ \ \Bigl. + \ (\epsilon_1 \, k_3) \, (\epsilon_2 \, k_4) \,
(k_2^\mu \,
\al_{\mu \nu} \, \epsilon^\nu_3) \ - \ (\epsilon_3 \, k_1) \, (\epsilon_2 \,
k_4) \, (k_2^\mu
\, \al_{\mu \nu} \, \epsilon^\nu_1) \Bigr. \notag \\
& \ \ \ \ \ \ \ \Bigl. + \ \frac{1}{2\al'} \; (\epsilon_3 \, k_1) \,
(\epsilon^\mu_1 \,
\al_{\mu \nu} \, \epsilon^\nu_2) \ - \ \frac{1}{2\al'} \; (\epsilon_1 \, k_3) \,
(\epsilon^\mu_2 \, \al_{\mu \nu} \, \epsilon^\nu_3) \Bigr. \notag \\
& \ \ \ \ \ \ \ \Bigl. - \ \frac{u}{2} \; (\epsilon_1 \, \epsilon_3) \, (k_1^\mu
\,
\al_{\mu \nu} \, \epsilon_2^\nu) \ + \ \frac{s}{2} \; (\epsilon_1 \, \epsilon_3)
\,
(k_3^\mu \, \al_{\mu \nu} \, \epsilon_2^\nu) \, \Bigr] \notag \\
& \ \ \ \biggl. \ - \ \frac{1}{2} \; \Bigl[ \, (\epsilon_1 \, \epsilon_2) \,
\epsilon_3^\mu
\, \al_{\mu \nu} \, (k_2^\nu - k_1^\nu) \ + \  (\epsilon_2 \, \epsilon_3) \,
\epsilon_1^\mu \,
\al_{\mu \nu} \, (k_3^\nu - k_2^\nu) \ + \  (\epsilon_1 \, \epsilon_3) \,
\epsilon_2^\mu \,
\al_{\mu \nu} \, (k_1^\nu - k_3^\nu) \, \Bigr] \biggr\} \ ,
\label{4pta}
\end{align}
with the color factor
\begin{eqnarray}
t^{a_1a_2a_3a_4} &=& \nonumber
{\rm Tr}(T^{a_1}T^{a_2}T^{a_3}T^{a_4})-{\rm Tr}(T^{a_4}T^{a_3}T^{a_2}T^{a_1})\\
&=& i \, ( d^{a_1a_4n}f^{a_2a_3n} - d^{a_2a_3n}f^{a_1a_4n})\label{tfact}\ ,
\end{eqnarray}
{\it i.e}.\ $t^{a_1a_2a_3a_4}=\Tr(T^{a_1}T^{a_2}T^{a_3}T^{a_4})-
\Tr(T^{a_1}T^{a_4}T^{a_3}T^{a_2})$,\
$t^{a_2a_3a_1a_4}=\Tr(T^{a_1}T^{a_4}T^{a_2}T^{a_3})-\Tr(T^{a_1}T^{a_3}T^{a_2}T^{
a_4})$ and
$t^{a_3a_1a_2a_4}=\Tr(T^{a_1}T^{a_2}T^{a_4}T^{a_3})-\Tr(T^{a_1}T^{a_3}T^{a_4}T^{
a_2})$.
Note that the massless \req{FinalIV} and massive \req{4pta} amplitudes
have different group structures, {\it c.f}.\
 Eq.\req{tfact} and
\req{Tfact}, respectively. This is explained in the following.

Generally, under world--sheet parity an $N$--point open superstring
amplitude\footnote{Recall the definition \req{BASIC}.} $A(1,\ldots,N)$ behaves
as
\be\label{worldsheetparity}
A(1,\ldots,N)=\lf(\prod_{i=1}^N (-1)^{\ap m_i^2+\eps}\ri)\ A(N,\ldots,1)\ ,
\ee
with $m_i^2$ the masses of the external open string states. Furthermore, for
$SO(N)$ representations we have $\eps=1$ and $\eps=0$ for $USp(N)$
representations
\cite{Green:1987sp,Angelantonj:2002ct}.
Further relations between subamplitudes are obtained by analyzing
their monodromy behavior w.r.t. to contour integrals in the complex plane
\cite{Stieberger:2009hq}.
As a consequence for amplitudes involving only massless external string
states ($m_i^2=0$)
the full set of relations allows to reduce the number of independent
subamplitudes to $(N-3)!$ \cite{Stieberger:2009hq,BjerrumBohr:2009rd}.
However, the set of relations for the massless case does not
hold in the case if $m_i^2\neq 0$ and new
monodromy relations have to be derived.

For the case at hand, {\it i.e}.\ $m_i=0,\ i=1,2,3$ and $m_4^2=\ap^{-1}$,
the partial amplitudes are odd under the parity transformation. Hence from
\req{worldsheetparity} we deduce:
\bea\label{odd}
A(1,2,3,4)&=&-A(1,4,3,2)\ ,\\
A(1,2,4,3)&=&-A(1,3,4,2)\ ,\\
A(1,3,2,4)&=&-A(1,4,2,3)\ .
\eea
This fact is manifest in the full amplitude \req{4pta} due to the color factor \req{tfact}.
After applying the contour arguments of \cite{Stieberger:2009hq}
the following monodromy relation can be established
for the case at hand:
\be\label{monodromie}
A(1,2,3,4)-e^{i\pi \ap s}\ A(1,2,4,3)-e^{-i\pi \ap u}\ A(1,4,2,3)=0\ .
\ee
Together with \req{odd} this relation allows to express all
six partial amplitudes in terms of one, say $A(1,2,3,4)$:
\bea\label{allow}
\ds{A(1,4,3,2)}&=&\ds{-A(1,2,3,4)\ ,}\\[5mm]
\ds{A(1,2,4,3)}&=&\ds{-A(1,3,4,2)=
\fc{\sin(\pi \ap u)}{\sin(\pi \ap t)}\ A(1,2,3,4)\ ,}\\[5mm]
\ds{A(1,3,2,4)}&=&\ds{-A(1,4,2,3)=-\fc{\sin(\pi \ap s)}{\sin(\pi \ap t)}\
A(1,2,3,4)\ .}
\eea
Note, that \req{monodromie} differs from the monodromy relation
for the massless case, c.f. Eq. (4.8) of \cite{Stieberger:2009hq}.
As a consequence also the solution \req{allow} is different than in the massless
case, c.f. Eq. (4.10) of \cite{Stieberger:2009hq}.
It is easy to see that the relations \req{allow}
are indeed  satisfied by the result \req{4pta}.

In order to represent the  amplitude
\req{4pta} in the helicity basis, we rewrite it as:
\begin{equation}
B[\alpha;\epsilon_1,\epsilon_2,\epsilon_3] ~=~
8\, g^2 \big( \, V_t \, t^{a_1a_2a_3a_4}  +  V_s \,
t^{a_2a_3a_1a_4}
 +  V_u \, t^{a_3a_1a_2a_4} \,\big)\
\sqrt{2\al'}\ \mathscr{B}[\alpha;\epsilon_1,\epsilon_2,\epsilon_3] \ .
\label{b3g}\end{equation}
We find
\begin{equation}
 \mathscr{B}[\alpha;\pm,\pm,\pm]=0\ ,
\end{equation}
therefore non-vanishing amplitudes always involve one gluon of a given helicity
and two of the opposite one. They have a very simple form:
\begin{flalign}
\mathscr{B}\left[-2;+,+,-\right] &
~=~\frac{1}{2\sqrt{2}}\ \!\frac{\langle
p3\rangle^{4}}{\langle12\rangle\langle23\rangle
\langle31\rangle}\ ,\nonumber\\[1mm]
\mathscr{B}\left[-1;+,+,-\right] &
~=~\frac{1}{\sqrt{2}}\ \,\frac{\langle
p3\rangle^{3}\langle3q\rangle}{
\langle12\rangle\langle23\rangle\langle31\rangle}\ ,\nonumber\\[1mm]
\mathscr{B}\left[~0~;+,+,-\right] &
~=~\frac{\sqrt{3}}{2}\ \,\frac{\langle
p3\rangle^{2}\langle3q\rangle^{2}}{
\langle12\rangle\langle23\rangle\langle31\rangle}\ ,\label{b3gh}\\[1mm]
\mathscr{B}\left[+1;+,+,-\right] &
~=~\frac{1}{\sqrt{2}}\ \,\frac{\langle
q3\rangle^{3}\langle3p\rangle}{
\langle12\rangle\langle23\rangle\langle31\rangle}\ ,\nonumber\\[1mm]
\mathscr{B}\left[+2;+,+,-\right] &
~=~\frac{1}{2\sqrt{2}}\ \!\frac{\langle
q3\rangle^{4}}{\langle12\rangle\langle23\rangle\langle31\rangle}\ .
\nonumber\end{flalign}

Similarly to the two-gluon case, we can consider the case of unpolarized $B$
decaying into a specific
helicity configuration of the three gluons. By using Eqs. (\ref{b3g}) and
(\ref{b3gh}), we obtain
\begin{equation}
 \sum_{\alpha=-2}^{+2}|B(\alpha;+,+,-)|^2=16\, g^4\,\frac{(1-\al'
s)^4}{\al^{\prime 3}\,s\,t\, u}
\,| V_t \, t^{a_1a_2a_3a_4}  +  V_s \,
t^{a_2a_3a_1a_4}
 +  V_u \, t^{a_3a_1a_2a_4}|^2\ .\label{prob3g}
\end{equation}

Next, we turn to $B$-decays into fermions.
The quark-antiquark channel is described by
the amplitude
\begin{equation}
B[\alpha;u_1,\bar u_2]=(T^a)^{\alpha_1}_{\alpha_2}(g\sqrt{2\al'})\,
k_1^{\mu}\alpha_ { \mu\nu }\, u_1^{\lambda}\sigma^{\nu}_{\lambda\dot\rho}\bar
u^{\dot\rho}_2\ ,\label{bqq}\end{equation}
which we rewrite as:
\begin{equation}
 B[\alpha;u_1,\bar u_2]=g\,(T^a)^{\alpha_1}_{\alpha_2}\ (2\al')^{3/2}\mathscr{B}
\left[
\alpha;u_1,\bar u_2\right]\ .
\end{equation}
{}For the specific $(\hpl,\hmi)$ helicity configuration of the antiquark-quark
pair, we obtain:
\begin{eqnarray}
\mathscr{B}\left[-2;\hpl,\hmi\right] &
=&\frac{1}{2}\ \langle p1\rangle\langle p2\rangle[q1]^{2}\ ,
\nonumber\\
\mathscr{B}\left[-1;\hpl,\hmi\right] &
=& \frac{1}{4}\ \langle p2\rangle[q1]\,\big(\langle q1\rangle[1q]-3\langle
p1\rangle[1p]\big)\ ,    \nonumber\\
\mathscr{B}\left[~0~;\hpl,\hmi\right] & =& \frac{\sqrt{6}}{4}\,\langle
2p\rangle[p1]\,\big(\langle p1\rangle[1p]-\langle q1\rangle[1q]\big)\ ,\\
\mathscr{B}\left[+1;\hpl,\hmi\right] & =
&\frac{1}{4}\ \langle q2\rangle[p1]\,\big(\langle p1\rangle[1p]-3\langle
q1\rangle[1q]\big)\ ,  \nonumber\\
\mathscr{B}\left[+2;\hpl,\hmi\right] & =
&  \frac{1}{2}\ \langle q1\rangle\langle q2\rangle[p1]^{2}\ .\nonumber
\end{eqnarray}
Adding up the moduli squares of the amplitudes, we obtain
\begin{equation}
 \sum_{\alpha=-2}^{+2}|B(\alpha;\hpl,\hmi)|^2=\frac{1}{2\al'}\, g^2
\,[(T^a)^{\alpha_1}_{\alpha_2}]^2\ ,\label{prob2q}
\end{equation}
which does not depend on the choice of the reference vectors. As a further
check, we can compare our result with the residue of the two-gluon --
quark-antiquark amplitude
\begin{equation}
{\rm Res}_{s=1/\ap}\ {\cal M}[q_1^-,\bar q_2^+,g_3^-,g_4^+] \eq  2\,g^2
\{T^{a_3}T^{a_4}\}^{\alpha_1}_{\alpha_2}\ \al' tu\, \frac{\langle
13\rangle^2}{\langle 14\rangle\langle 24\rangle}\ , \label{qqfact}
\end{equation}
which is known to receive contributions from the spin 2 resonance only
\cite{Anchordoqui:2008hi}.
Indeed, the residue is correctly reproduced by
\begin{equation}
  \sum_{\alpha=-2}^{+2}B(\alpha;\hpl,\hmi)^*\, B(\alpha;-,+)|_{(1\to 3,2\to 4)}\
.\label{sumb1}
\end{equation}

The amplitude with one gluon in addition to the quark-antiquark pair in the
final state can be written as:
\begin{equation}
 B[\alpha;u_1,\bar
u_2,\epsilon_3]=2\,g^2\ \big[V_t(T^{a_3}T^{a_4})^{\alpha_1}_{\alpha_2}+
 V_u(T^{a_4}T^{a_3})^{\alpha_1}_{\alpha_2}\big]\ \sqrt{2\al'}\ \mathscr{B}
\left[
\alpha;u_1,\bar u_2,\epsilon_3\right]\ ,\label{bqq1}
\end{equation}
with:
\begin{eqnarray}
\mathscr{B}\left[-2;\hpl,\hmi,+\right] &
=&\frac{1}{\sqrt{2}}\ \frac{\langle p1\rangle\langle
p2\rangle^3}{\langle12\rangle\langle23\rangle
\langle31\rangle}\ ,
\nonumber\\[1mm]
\mathscr{B}\left[-1;\hpl,\hmi,+\right] &
=& \frac{1}{2\sqrt{2}}\,
\frac{\langle p2\rangle^2}{\langle12\rangle\langle23\rangle
\langle31\rangle}
\big(\langle q1\rangle\langle p2\rangle +3\langle p1\rangle\langle
q2\rangle\big)\ ,
\nonumber\\[1mm]
\mathscr{B}\left[~0~;\hpl,\hmi,+\right] & =& \frac{\sqrt{3}}{2}\
\frac{\langle p2\rangle\langle q2\rangle}{\langle12\rangle\langle23\rangle
\langle31\rangle}\ \big(\langle q1\rangle\langle p2\rangle +\langle
p1\rangle\langle q2\rangle\big)\ ,\label{bqq2}\\[1mm]
\mathscr{B}\left[+1;\hpl,\hmi,+\right] & =
&\frac{1}{2\sqrt{2}}\,
\frac{\langle q2\rangle^2}{\langle12\rangle\langle23\rangle
\langle31\rangle}\big(3\langle q1\rangle\langle p2\rangle +\langle
p1\rangle\langle q2\rangle\big) \ , \nonumber\\[1mm]
\mathscr{B}\left[+2;\hpl,\hmi,+\right] & =
&  \frac{1}{\sqrt{2}}\
\frac{\langle q1\rangle\langle q2\rangle^3}{\langle12\rangle\langle23\rangle
\langle31\rangle}\ .\nonumber
\end{eqnarray}
{}For the gluon with opposite helicity we have:
\begin{equation}
\mathscr{B}\left[\al;\hpl,\hmi,-\right]
=\mathscr{B}^*\!\left[\al;\hpl,\hmi,+\right]\Big|_{(p\leftrightarrow
q),(1\leftrightarrow 2)}\ .
\end{equation}
The sum of the squared moduli of the corresponding amplitudes reads
\begin{eqnarray}
 \sum_{\alpha=-2}^{+2}|B(\alpha;\hpl,\hmi,+)|^2 &=&  g^4
 \big|V_t(T^{a_3}T^{a_4})^{\alpha_1}_{\alpha_2}+
 V_u(T^{a_4}T^{a_3})^{\alpha_1}_{\alpha_2}\big|^2\nonumber \\ &&\times
 \frac{(1-\al't)^2}{{\al'}^2s\,t\,u}\ (s+4{\al'}tu)\
,\label{prob2qg}
\end{eqnarray}
and a similar expression with $(k_1\leftrightarrow k_2, a_3\leftrightarrow a_4)$
for the gluon with opposite helicity.

\subsection{Massive spin 1 boson $W(J=1)$}

\noindent
The spin one vector resonance has a different character than spin two, because
it is
tied to space--time SUSY. The internal part of the corresponding vertex
operator (\ref{vertw}) contains the current $\cal J$, which plays an important
role in the world--sheet SCFT  describing superstrings propagating on CYMs.
As we have described in Section III the most natural way of thinking about
this particle is as a two--gluino bound state.
Indeed, with  $\Sigma$  the internal Ramond
field associated to the gluino, c.f. Eq. (\ref{vertlam}), the current $\Jc$
appears as a sub-leading term in the OPE \req{OPE1}.
It is clear that the current $\Jc$ and the existence of the resonance $W(J=1)$
is a universal property of all $\Nc=1$ SUSY compactifications.
At the disk level, this particle does not couple to purely gluonic processes.
Its main decay channel is into two gluinos and its mass will be affected by the
SUSY breaking mechanism.
The reason why we include it in our discussion is that it also couples to the
quark sector, therefore it can be {\it a priori}\/ directly produced at the LHC.

In the intersecting D--brane models, the internal part of the quark vertex
operators \req{QUARKV} contains the boundary-changing operators
\cite{Lust:2008qc}
\begin{equation}
\Xi^{a \cap b} (z) \eq \prod_{j=1}^3 e^{i \, (\frac{1}{2}-\theta^j) \, H_j(z)}
\, \si_{\theta^j}(z) \co \bar \Xi^{a \cap b} (z) \eq \prod_{j=1}^3 e^{-i \,
(\frac{1}{2}-\theta^j) \, H_j(z)} \, \si_{-\theta^j}(z)\ ,
\label{univ4}
\end{equation}
where $\sigma_{\theta}$ is the bosonic twist operator associated to the
intersection angle $\theta$.
The angles $\theta^i$ are associated to the three complex planes  subject to the
$\Nc=1$ SUSY constraint:
\begin{equation}
\sum_{k=1}^3\theta^k=0\ .
\end{equation}
Note, that in the limit $\theta\ra0$ bosonic twist fields $\si_\th$ become the
identity operator  and we have:
\beq
\lim_{\theta^j \rightarrow 0} \Xi^{a \cap b}  \eq \prod_{j=1}^3 e^{ \frac{i}{2}
\, H_j} \eq \Si \co \lim_{\theta^j \rightarrow 0} \bar \Xi^{a \cap b} \eq
\prod_{j=1}^3 e^{-\frac{i}{2} \, H_j} \eq \bar \Si \ .
\label{univ11}
\end{equation}
Therefore, up to Chan-Paton and normalization factors in this limit
the quark vertex operators \req{QUARKV} turn into the gaugino vertex operators
\req{vertlam}.
With the  explicit free field representation of the $U(1)$ current $\Jc$
\be\label{univ2}
\Jc=i\sqrt 3\ \p H=i\ \sum_{j=1}^3 \pa H_j\ ,
\ee
the three-point function relevant to the $W$ coupling to a quark-antiquark pair
reads:
\begin{eqnarray}
\langle {\cal J}(z_1) \, \Xi^{a \cap b} (z_2) \, \bar  \Xi^{a \cap b} (z_3)
\rangle &=& \sum_{j=1}^3 \bigl(  \tfrac{1}{2} \, - \, \theta^j \bigr) \;
\frac{z_{23}^{1/4} }{z_{12} \, z_{13}} \eq \frac{3 \, z_{23}^{1/4} }{2 \, z_{12}
\, z_{13}}\nonumber \\[2mm] &=&
\langle {\cal J}(z_1)  \Si (z_2) \ \bar  \Si (z_3)\rangle\ .
\label{univ12}
\end{eqnarray}
The corresponding amplitude is:
\begin{equation}
W[\xi,u_1,\bar u_2] ~=~\frac{\sqrt{3}}{4}\ g(T^a)^{\alpha_1}_{\alpha_2}\
\xi_{\mu}u_1^{\lambda}\sigma^{\mu}_{\lambda\dot\rho}\bar u^{\dot\rho}_2~\equiv~
\sqrt{\frac{3\al'}{2}}\ g(T^a)^{\alpha_1}_{\alpha_2}\
\mathscr{W}[\xi,u_1,\bar u_2]\ .
\label{wamp}
\end{equation}
{}For the specific $(\hpl,\hmi)$ helicity configuration of the antiquark-quark
pair, we obtain
\begin{eqnarray}
\mathscr{W}\left[-1;\hpl,\hmi\right] &
=& \ \langle p2\rangle[q1] \ ,   \nonumber\\
\mathscr{W}\left[~0~;\hpl,\hmi\right] & =& \sqrt{2}\langle p2\rangle[p1]\ ,\\
\mathscr{W}\left[+1;\hpl,\hmi\right] & =
&\  \langle q2\rangle[p1]  \ .\nonumber
\end{eqnarray}
{}From Eq. (\ref{univ12}) it follows that the $W$-coupling to two gauginos can
be
obtained from Eq. (\ref{wamp}) by the replacement
$(T^a)^{\alpha_1}_{\alpha_2}\to
4 d^{a_1a_2a}$.
The normalization of the above couplings can be checked by comparing with
Eq. (39) of Ref. \cite{Anchordoqui:2008hi}.

The amplitude with one gluon in addition to the quark-antiquark pair in the
final state can be written as:
\begin{equation}
 W[\xi;u_1,\bar
u_2,\epsilon_3]=\sqrt{3}\,g^2\ \big[V_t(T^{a_3}T^{a_4})^{\alpha_1}_{\alpha_2}+
 V_u(T^{a_4}T^{a_3})^{\alpha_1}_{\alpha_2}\big]\ \mathscr{W} \left[
\xi;u_1,\bar u_2,\epsilon_3\right]\ .
\end{equation}
with
\begin{eqnarray}
\mathscr{W}\left[-1;\hpl,\hmi,+\right] &
=&
~\frac{\langle p2\rangle^2}{\langle13\rangle\langle23\rangle}\ ,
\nonumber\\[1mm]
\mathscr{W}\left[~0~;\hpl,\hmi,+\right] & =& \sqrt{2}\,
\frac{\langle p2\rangle\langle q2\rangle}{\langle13\rangle\langle23\rangle
}\ ,\\[1mm]
\mathscr{W}\left[+1;\hpl,\hmi,+\right] & =
&
-\frac{\langle q2\rangle^2}{\langle13\rangle\langle23\rangle}\ . \nonumber
\end{eqnarray}
{}For the gluon with opposite helicity we have:
\begin{equation}
\mathscr{W}\left[\xi;\hpl,\hmi,-\right]
=\mathscr{W}^*\!\left[\xi;\hpl,\hmi,+\right]\Big|_{(p\leftrightarrow
q),(1\leftrightarrow 2)}\ .
\end{equation}
The sum of the squared moduli of the corresponding amplitudes reads
\begin{equation}
 \sum_{\xi=-1}^{+1}|W(\xi;\hpl,\hmi,+)|^2 = 3\, g^4
 \big|V_t(T^{a_3}T^{a_4})^{\alpha_1}_{\alpha_2}+
 V_u(T^{a_4}T^{a_3})^{\alpha_1}_{\alpha_2}\big|^2\
 \frac{(1-\al't)^2}{{\al'}^2\,t\,u}\label{wprobqg}
\end{equation}
and a similar expression with $(k_1\leftrightarrow k_2, a_3\leftrightarrow a_4)$
for the gluon with opposite helicity.

\subsection{The universal scalar $\Phi(J=0)$}

\noindent
It has been originally pointed out in Ref. \cite{Anchordoqui:2008hi} that the
lowest scalar resonance propagating in two-particle channels of multi-gluon
amplitudes must couple to  the product of ``self-dual'' gauge field strengths,
with the coupling to two gluons that is non-vanishing only if they carry the
same helicities, say $(+,+)$. Such couplings arise naturally from $\Nc=1$
supersymmetric $F$-terms
 $\int d^2\theta\,\Phi W^{\alpha}W_{\alpha}$ where $W^{\al}$ is the gauge field
strength superfield. The scalar and pseudoscalar components of complex
$\Phi\equiv\Phi_+$ $(\Phi_-=\bar\Phi)$ are combined with the relative weight
that enforces this selection rule.

The two-gluon decay of $\Phi$ with momentum $k$ is described by the amplitude
\begin{eqnarray}
\Phi_{\pm}[\epsilon_1,\epsilon_2]&=&2g\ (2d^{a_1a_2a})\ \sqrt{2\alpha'}\
\Big\{\
(g_{\mu\nu}
+2\alpha'k_{\mu}k_{\nu})\big[(\epsilon_{2}
k_{1})k_{2}^{\mu}\epsilon_{1}^{\nu}+(\epsilon_{1}
k_{2})k_{1}^{\mu}\epsilon_{2}^{\nu}\nonumber\\ &&-(k_{1}
k_{2})\epsilon_{2}^{\mu}\epsilon_{1}^{\nu}-(\epsilon_{1}
\epsilon_{2})k_{1}^{\mu}k_{2}^{\nu}\big]
\pm i \epsilon_{\mu\nu\rho\lambda}k^{\lambda}
\epsilon_{1}^{\mu}\epsilon_{2}^{\nu}k_{2}^{\rho}\ \Big\}\ .\label{VERT}
\end{eqnarray}
Note, that the last piece of the
vertex \req{VERT} is reminiscent to the generic vertex \req{Massive1}.
In the helicity basis,
\beq
\Phi_+[-,-]=\Phi_+[-,+]=\Phi_+[+,-]=0\ ,
\end{equation}
and
\beq
\Phi_+[+,+]~=~2g(2d^{a_1a_2a})\sqrt{2\alpha'}\,[12]^2\ .
\end{equation}
The conjugate scalar $\Phi_-$ couples to $(-,-)$ configuration only, with the
complex conjugate coupling. Our results correctly reproduce Eq. (25) of
Ref. \cite{Anchordoqui:2008hi}.

The three-gluon decay amplitudes obey similar selection rules:
\begin{equation}
\Phi_+[-,-,-]=\Phi_+[-,-,+]=\Phi_+[-,+,-]=\Phi_+[+,-,-]=0\ ,
\end{equation}
while the non-vanishing ones are the ``all plus'' amplitude
\begin{equation}
\Phi_+[+,+,+]
 ~=~
4\, g^2 \big( \, V_t \, t^{a_1a_2a_3a_4}  +  V_s \,
t^{a_2a_3a_1a_4}
 +  V_u \, t^{a_3a_1a_2a_4} \,\big)
\frac{(\al')^{-3/2}}{\langle 12\rangle \langle 23\rangle  \langle 31\rangle}
\label{f3g1}\end{equation}
and three ``mostly plus'' amplitudes that can be obtained from
\begin{equation}
\Phi_+[+,+,-]
 ~=~
4\, g^2 \big( \, V_t \, t^{a_1a_2a_3a_4}  +  V_s \,
t^{a_2a_3a_1a_4}
 +  V_u \, t^{a_3a_1a_2a_4} \,\big)\sqrt{\al'}
\frac{[12]^4}{[12][23][31]} \ ,
\label{f3g2}\end{equation}
by cyclically permuting $(1,2,3)$.

The $\Phi$ resonance couples to the quark-antiquark pair and one gluon only if
the gluon is in appropriate polarization state: $+$ for $\Phi^+$ and $-$ for
$\Phi^-$. The amplitude reads
\begin{equation}
 \Phi_+[\hpl,\hmi,+]=2g^2\big[V_t(T^{a_3}T^{a_4})^{\alpha_1}_{
\alpha_2 } +
 V_u(T^{a_4}T^{a_3})^{\alpha_1}_{\alpha_2}\big]\,\sqrt{\al'}\,\frac{[13]^2}{[12]
 } \ .
\end{equation}

\subsection{The Calabi-Yau scalar $\Omega(J=0)$}

\noindent
The universal $\Omega$ scalar in (\ref{vertom})
is  associated to world--sheet operator $\Oc$ appearing in the $\Nc=1$ OPEs
\req{OPE1}.
The field $\Oc$ comprising the internal part of the vertex operator
(\ref{vertom})
has charge $3$ w.r.t. the internal current $\Jc$.
Hence the field $\Omega(J=0)$ does not couple to purely gluonic processes at the
disk level, similarly to
$W(J=1)$. It can couple though to two fermions of the same helicity. The
coupling to two quarks (color triplets) is not allowed because $\Omega$ is a
color octet, but the coupling to two gluinos is non-vanishing and can be used to
determine the normalization factor of the respective vertex operator. The LHC
production rate of this particle is suppressed at least by ${\cal
O}(\alpha_s^2)$ compared to other resonances, therefore we do not discuss it
here any further.

\subsection{Massive spin 3/2 quark $Q^{\star}(J=3/2)$}

\noindent
Massive quarks are color triplets [in general, in the fundamental representation
of $U(N)$]. Their main decay channels are into a quark and a gluon. {}For the
spin 3/2 resonance\linebreak $Q^{\star}(J=3/2)$, the respective amplitude reads
\begin{eqnarray}
Q^{\star}(R;\epsilon_1,u_2)&=&(T^{a_1})_{\alpha}^{\al_2}g\,\sqrt{2\al'}
(k_{1\mu}\epsilon_{1\nu}-k_{1\nu}\epsilon_{1\mu})\,
u_2^{\lambda}\sigma^{\mu}_{\lambda\dot
\rho}\bar\chi^{\nu\dot\rho}\nonumber\\[1mm]
&\equiv&(T^{a_1})_{\alpha}^{\al_2}\sqrt{2}g\,\al'
\mathscr{Q}^{\star}
(R;\epsilon_1,u_2)\ ,
\end{eqnarray}
where $\bar\chi$ is the lower component of the vector spinor $R$,
see Eqs. (\ref{umu1}) and (\ref{umu2}).
In the helicity basis,
\begin{equation}
\mathscr{Q}^{\star}(R;+,\hpl)=0\label{selq}
\end{equation}
and:
\begin{flalign}
\mathscr{Q}^{\star}(\tmi;-,\hpl)&
=\langle p1\rangle^{2}[2q]\ ,\\
\mathscr{Q}^{\star}(\hmi;-,\hpl)&
=\sqrt{3}\langle p1\rangle\langle q1\rangle[q2]\ ,\\
\mathscr{Q}^{\star}(\hpl;-,\hpl)&
=\sqrt{3}\langle p1\rangle\langle q1\rangle[p2]\ ,\\
 \mathscr{Q}^{\star}(\tpl;-,\hpl)&
=\langle q1\rangle^{2}[2p]\ .\end{flalign}
The above result agrees with Eq. (47) of
Ref. \cite{Anchordoqui:2008hi}.
Adding up the moduli squares of the  amplitudes, we obtain:
\begin{equation}
\sum_{R=\tmi}^{\tpl}|\mathscr{Q}^{\star}(R;-,\hpl)|^{2}=g^2\,
|(T^{a_1})_{\alpha}^{\al_2}|^2\frac{2}{\al'}\ .
\end{equation}

The amplitude with one quark and two gluons in the final state reads:
\begin{eqnarray}
Q^{\star}(R;\epsilon_1 &&,\epsilon_2,u_3) ~=~ 2\,g^2\sqrt{2\al'}\,
[ \, V_t \, (T^{a_1}  T^{a_2})^{\al_3} _{\al_4} \
- \ V_u \, (T^{a_2}  T^{a_1})_{\al_4}^{\al_3}\, ]  \notag \\
&& \!\times \bigg\{ \, \frac{1}{s} \; \Big[ \, (\epsilon_2 \, k_1) \, k_2^\mu \,
(\bar \chi_\mu \! \not \! \epsilon_1 u_3) \ - \ (\epsilon_1 \, k_2) \, k_1^\mu
\, (\bar
\chi_\mu \! \not \! \epsilon_2 u_3) \ + \ (\epsilon_2 \, k_1) \, k_1^\mu \,
(\bar \chi_\mu
\! \not \! \epsilon_1 u_3)\notag \\[1mm]
&& \qquad - \ (\epsilon_1 \, k_2) \, k_2^\mu \, (\bar
\chi_\mu \! \not \!
\epsilon_2 u_3) +  \ (\epsilon_1 \, \epsilon_2) \, k_1^\mu \, (\bar \chi_\mu
\! \not \! k_2 u_3) \ - \ (\epsilon_1 \, \epsilon_2) \, k_2^\mu \, (\bar
\chi_\mu \! \not \!
k_1 u_3) \notag\\[1mm]
&& \qquad - \ (\epsilon_1 \, k_2) \, \epsilon_2^\mu \, (\bar \chi_\mu \!
\not \! k_4 u_3) \
+ \ ( \epsilon_2 \, k_1) \, \epsilon_1^\mu \, (\bar \chi_\mu \! \not \! k_4 u_3)
\, \Bigr]
\notag \\
&& \quad\, + \,\frac{1}{t} \; \Big[ \, (\epsilon_1 \, k_3) \, k_2^\mu \, (\bar
\chi_\mu \! \not \! \epsilon_2 u_3) \ - \ (\epsilon_1 \, k_3) \, \epsilon_2^\mu
\, (\bar \chi_\mu
\! \not \! k_2 u_3)  \notag \\
&& \qquad\qquad - \ \frac{1}{2} \, \epsilon_2^\mu \,
(\bar \chi_\mu \! \not \! k_2 \! \not \! \epsilon_1 \! \not \! k_1 u_3) \ - \
\frac{1}{2} \, k_2^\mu \, (\bar \chi_\mu \! \not \! \epsilon_2 \! \not \! k_1 \!
\not
\! \epsilon_1 u_3) \, \Big] \notag \nonumber
\end{eqnarray}
\begin{eqnarray}
 && \quad\, +\, \ \frac{1}{u} \; \Big[ \, (\epsilon_2 \, k_3) \, \epsilon_1^\mu
\,
(\bar \chi_\mu \! \not \! k_1 u_3) \ - \ (\epsilon_2 \, k_3) \, k_1^\mu \, (\bar
\chi_\mu \! \not \! \epsilon_1 u_3)  \notag \\
 &&\qquad\qquad + \ \frac{1}{2} \, k_1^\mu \, (\bar
\chi_\mu \! \not \! \epsilon_1 \! \not \! k_2 \! \not \! \epsilon_2 u_3) \ + \
\frac{1}{2}
\, \epsilon_1^\mu \, (\bar \chi_\mu \! \not \! k_1 \! \not \! \epsilon_2 \! \not
\! k_2 u_3)
\, \Big] \notag \\
 && \qquad - \ \frac{1}{2} \; \epsilon_1^\mu \, (\bar \chi_\mu
\! \not \! \epsilon_2 u_3)  \ + \ \frac{1}{2} \; \epsilon_2^\mu \, (\bar
\chi_\mu \!
\not \! \epsilon_1 u_3) \, \bigg\} \ .
\label{4ptg}
\end{eqnarray}
It is convenient to rewrite this amplitude as:
\begin{equation}
Q^{\star}(R;\epsilon_1,\epsilon_2,u_3) ~=~ 2\,g^2\,
[ \, V_t \, (T^{a_1}  T^{a_2})^{\al_3} _{\al_4} \
- \ V_u \, (T^{a_2}  T^{a_1})_{\al_4}^{\al_3}\, ]\,
\mathscr{Q}^{\star}(R;\epsilon_1,\epsilon_2,u_3)\ .
\end{equation}
We find the selection rule:
\begin{equation}
 \mathscr{Q}^{\star}(R;+,+,\hpl)=0\ .\label{sel2}
\end{equation}
{}For two gluons in the $(-,-)$ helicity configuration, the amplitude reads:
\begin{flalign}
\mathscr{Q}^{\star}(\tmi;-,-,\hpl)&
=~\frac{[3q]^{3}}{[12][23][31]}\ ,\notag\\
\mathscr{Q}^{\star}(\hmi;-,-,\hpl)&
=\sqrt{3}\frac{[3q]^{2}[p3]}{[12][23][31]}\ ,\notag\\
\mathscr{Q}^{\star}(\hpl;-,-,\hpl)&
=\sqrt{3}\frac{[3p]^{2}[q3]}{[12][23][31]}\ ,\\
\mathscr{Q}^{\star}(\tpl;-,-,\hpl)&
=~\frac{[3p]^{3}}{[12][23][31]}\notag\ .\end{flalign}
When the gluons carry opposite helicities, then:
\begin{flalign}
\mathscr{Q}^{\star}(\tmi;+,-,\hpl)&
=~~\sqrt{\al'}\frac{\langle p2\rangle^{3}}{\langle 12\rangle\langle
13\rangle}\ ,\notag\\
\mathscr{Q}^{\star}(\hmi;+,-,\hpl)&
=~\sqrt{3\al'}\;\frac{\langle p2\rangle^{2}\langle q2\rangle}{\langle
12\rangle\langle 13\rangle}\ ,\notag\\
\mathscr{Q}^{\star}(\hpl;+,-,\hpl)&
=-\sqrt{3\al'}\frac{\langle q2\rangle^{2}\langle p2\rangle}{\langle
12\rangle\langle 13\rangle}\ ,\\
\mathscr{Q}^{\star}(\tpl;+,-,\hpl)&
=~{-}\sqrt{\al'}\frac{\langle q2\rangle^{3}}{\langle 12\rangle\langle
13\rangle}\notag\ .\end{flalign}
and a similar expression with $(1\leftrightarrow 2)$ for gluons with flipped
helicities.

The sums of the squared moduli of the amplitudes read:
\begin{eqnarray}
\sum_{R=\tmi}^{\tpl}|Q^{\star}(R;-,-,\hpl)|^2 &=& 4\,g^4\,
|\, V_t \, (T^{a_1}  T^{a_2})^{\al_3} _{\al_4}
-  V_u \, (T^{a_2}  T^{a_1})_{\al_4}^{\al_3}\, |^2\frac{(1-\al' s)^3}{{\al'}^3
stu}\ ,\nonumber\\
\sum_{R=\tmi}^{\tpl}|Q^{\star}(R;+,-,\hpl)|^2 &=& 4\,g^4\,
|\, V_t \, (T^{a_1}  T^{a_2})^{\al_3} _{\al_4}
-  V_u \, (T^{a_2}  T^{a_1})_{\al_4}^{\al_3}\, |^2\frac{(1-\al' t)^3}{{\al'}^2
st}\ ,
\end{eqnarray}
\begin{eqnarray}
\sum_{R=\tmi}^{\tpl}|Q^{\star}(R;-,+,\hpl)|^2 &=& 4\,g^4\,
|\, V_t \, (T^{a_1}  T^{a_2})^{\al_3} _{\al_4}
-  V_u \, (T^{a_2}  T^{a_1})_{\al_4}^{\al_3}\, |^2\frac{(1-\al' u)^3}{{\al'}^2
su}\ .\nonumber
\end{eqnarray}

One important comment is here in order. Since in our conventions all particles
are incoming, the helicities of the final quark and gluons must be reversed in
the physical amplitudes describing decays of the excited quarks. Thus if the
$Q^{\star}$ fermion considered above decays into a number of gluons and only
one quark, the quark must be a left-handed $SU(2)$ doublet associated
to the intersection of the QCD and electro-weak branes [the $SU(2)$
index is just a spectator]. In order to produce a right-handed quark one would
have to start from another $Q^{\star}$ excitation, an $SU(2)$ singlet associated
to a different intersection of the QCD brane. Thus $Q^{\star}(J=3/2)$ [and
$Q(J=1/2$)] are the massive excitations of {\it chiral\/} fermions. In
superstring theory, there is no conventional ``doubling'' of massive
quarks because chiral fermions generate their own Regge trajectories.

Massive quark excitations can also decay into more  fermions. The
minimal case involves one quark and a fermion-antifermion pair in the final
state. The structure of the corresponding amplitudes is similar to four-fermion
processes discussed in Ref. \cite{Lust:2008qc}. Although lepton pairs can be
produced in this way, we focus on the case of two quarks and one antiquark, as
the most relevant to the direct production of $Q^{\star}$ and $Q$ in quark-quark
scattering and quark-antiquark annihilation at the LHC. Even in this case,
two qualitatively different computations need to be performed depending whether
the processes involve quarks form the intersection of the QCD brane with a
single brane (thus either four $SU(2)$ doublets or four $SU(2)$ singlets)
or from two intersections (amplitudes with both $SU(2)$ doublets and singlets).
In order to keep track of all gauge indices, it is convenient to display them
explicitly in the amplitudes. The lower $\alpha$ indices will label
$SU(3)$ triplets (stack $a$), the upper $\beta$ indices will label electroweak
$SU(2)$ doublets (stack $b$) and upper $\gamma$ (stack $c$) indices electroweak
singlets. Thus, for instance,
$Q^{\star}(R^{\beta}_{\al};{u}^{\alpha_1}_{\beta_1},\bar
u^{\beta_2}_{\alpha_2},u_{\beta_3}^{\alpha_3})$ will denote the amplitude with
the (incoming) $Q^{\star}$ Regge excitation of a left-handed quark, $\bar
q_{1R},q_{2L}$ and $\bar q_{3R}$. On the other hand,
$Q^{\star}(R^{\beta}_{\al}; \bar u^{\alpha_1}_{\gamma_1},
u^{\gamma_2}_{\alpha_2},u_{\beta_3}^{\alpha_3}  )$
will denote the amplitude with the same Regge excitation, $\bar
q_{1L}, q_{2R}$ and $\bar q_{3R}$.

We begin with the case of two stacks, say
$a$ and $b$, intersecting at angles
$\theta_j=\theta_{bj}-\theta_{aj},~ j=1,2,3$. By following the lines of
Ref. \cite{Lust:2008qc}, we obtain:
\begin{eqnarray}\nonumber
Q^{\star}&& ( R^{\beta}_{\al} ;{u}^{\alpha_1}_{\beta_1}, \bar
u^{\beta_2}_{\alpha_2},u_{\beta_3}^{\alpha_3})=
(2\al')^{3/2}e^{\phi_{10}} \int^1_0 \dd x\ I(x, \theta^j)\\[1mm]
&&\times\
\Big\{\delta_{\al}^{\alpha_1}\delta_{\alpha_2}^{\alpha_3}\delta_{\beta_1}^{
\beta_2 }
\delta_{\beta_3}^{\beta}\,Z^{ba}_{\rm
inst}(x)-\delta_{\al}^{\alpha_3}\delta_{
\alpha_2 } ^ { \alpha_1 } \delta_{\beta_3}^{\beta_2}
\delta_{\beta_1}^{\beta}\,Z^{ba}_{\rm
inst}(1-x)\Big\}\notag \\[1mm]
&& \times\
\Big\{ x^{-\al's} \, (1-x)^{-\al'u-1}\big[ \, (u_1
u_3) \, (\bar u_2  \bar \chi^\mu) \, k_\mu^1 \, + \, \tfrac{1}{4} \, (u_1 \!
\not \! k_4 \, \bar \chi^\mu) \, (u_3  \si_\mu \bar u_2)\ \label{4pti} \\
&&
\qquad\qquad + \, \tfrac{1}{4}\,
(u_3 \! \not \! k_4 \, \bar \chi^\mu) \, (u_1 \si_\mu \bar u_2) \, \big]
 \notag \\[1mm]
&& \qquad + \,x^{-\al's-1} \, (1-x)^{-\al'u}\big[ \, - \, (u_1 u_3) \,
(\bar u_2  \bar \chi^\mu) \, k_\mu^3 \, + \, \tfrac{1}{4} \, (u_1 \! \not \! k_4
\, \bar \chi^\mu) \, (u_3 \si_\mu \bar u_2)\nonumber \\
&& \qquad\qquad + \,
\tfrac{1}{4} \, (u_3 \! \not
\! k_4 \, \bar \chi^\mu) \, (u_1 \si_\mu \bar u_2) \, \big]
\Big\}\ .\nonumber
\end{eqnarray}
Here, $Z^{ba}_{\rm inst}$ is the instanton partition function
\cite{Lust:2008qc}. The function  $I(x, \theta^j)$, written explicitly in
Ref. \cite{Lust:2008qc}, is the correlation function of four
boundary-changing operators and it is symmetric under $x\to 1-x$.
It is convenient to define:
\begin{eqnarray}
Q_{su}&=&\al' e^{\phi_{10}} \int^1_0 \dd x\ Z^{ba}_{\rm
inst}(x)\ I(x, \theta^j)\ x^{-\al's} \, (1-x)^{-\al'u-1}\ .\nonumber \\
\widetilde{Q}_{su}&=&\al' e^{\phi_{10}} \int^1_0 \dd x\ Z^{ba}_{\rm
inst}(x)\ I(x, \theta^j)\ x^{-\al's-1} \, (1-x)^{-\al'u}\ .\label{qints}
\end{eqnarray}
Note that the amplitude  (\ref{4pti}) exhibits kinematical singularities
due to the propagation of massless gauge bosons in the respective
channels:
\begin{equation}
 Q_{su}\stackrel{u\to 0}{\longrightarrow} ~-\frac{g_b^2}{u}~,\qquad
\widetilde{Q}_{su}\stackrel{s\to 0}{\longrightarrow} ~-\frac{g_a^2}{s}\ .
\end{equation}
where $g_a=g$ and $g_b$ are  the QCD [more precisely $U(3)$] and electro-weak
coupling constants, respectively. In order to obtain the helicity amplitudes,
it is convenient to rewrite Eq. (\ref{4pti}) as
\begin{equation}Q^{\star}(R^{\beta}_{\al}
;{u}^{\alpha_1}_{\beta_1}, \bar
u^{\beta_2}_{\alpha_2},u_{\beta_3}^{\alpha_3})~=~
\delta_{\al}^{\alpha_1}\delta_{\alpha_2}^{\alpha_3}\delta_{\beta_1}^{
\beta_2 }\delta_{\beta_3}^{\beta}\,\al'\,\mathscr{Q}^{\star}( R
;{\hpl},\hmi,\hpl) ~-~ (1\leftrightarrow 3)\ ,\label{qq1}
\end{equation}
where:
\begin{flalign}
\mathscr{Q}^{\star}( \tmi
;{\hpl},\hmi,\hpl) &\eq Q_{su}\,\langle p2\rangle^{2}[q1][23]
~+~\widetilde{Q}_{su}\,\langle
p2\rangle^{2}[3q][12]\ ,\nonumber\\[1mm]
\mathscr{Q}^{\star}( \hmi
;{\hpl},\hmi,\hpl)&
\eq \frac{1}{\sqrt{3}}Q_{su}\,\langle p2\rangle[23]\Big\{2\langle
q2\rangle[q1]-\langle
p2\rangle[p1]\Big\}\nonumber\\ &\qquad\qquad +
~\frac{1}{\sqrt{3}}\widetilde{Q}_{su}\,\langle
p2\rangle[21]\Big\{2\langle q2\rangle[q3]-\langle
p2\rangle[p3]\Big\}\ ,\nonumber\\[1mm]
\mathscr{Q}^{\star}( \hpl
;{\hpl},\hmi,\hpl) &
\eq \frac{1}{\sqrt{3}}Q_{su}\,\langle q2\rangle[23]\Big\{2\langle
p2\rangle[p1]-\langle
q2\rangle[q1]\Big\}
\label{qqstar}\\ &\qquad\qquad + ~\frac{1}{\sqrt{3}}\widetilde{Q}_{su}\,\langle
q2\rangle[21]\Big\{2\langle p2\rangle[p3]-\langle
q2\rangle[q3]\Big\}\ ,\nonumber\\[1mm]
\mathscr{Q}^{\star}( \tpl
;{\hpl},\hmi,\hpl) &
\eq Q_{su}\,\langle q2\rangle^{2}[p1][23]~+ ~\widetilde{Q}_{su}\,\langle
q2\rangle^{2}[3p][12]\ .\nonumber\end{flalign}

{}Finally, we consider the case of three stacks, say
$a$, $b$ and $c$, intersecting at angles
$\theta_j=\theta_{bj}-\theta_{aj},~\nu_j=\theta_{cj}-\theta_{aj}, ~ j=1,2,3$.
Then the four-point correlation function of boundary-changing operators depends
on the additional set of angles: $I=I(x, \theta^j,\nu^j)$ \cite{Lust:2008qc},
however the rest of the computation is very similar to the two-stack case. Let
us define $R_{su}$ and $\widetilde{R}_{su}$ as the integrals (\ref{qints})
with $I(x, \theta^j)$ replaced by
$I(x, \theta^j,\nu^j)$ in the integrand, {\it i.e}.\ $Q_{su}\to  R_{su},~
\widetilde{Q}_{su}\to  \widetilde{R}_{su}$ upon  $I(x, \theta^j)\to
I(x, \theta^j,\nu^j)$. Then
the relevant amplitude can be written as
\begin{equation}
Q^{\star}(R^{\beta}_{\al}; \bar u^{\alpha_1}_{\gamma_1},
u^{\gamma_2}_{\alpha_2},u_{\beta_3}^{\alpha_3} )
~=~
\delta_{\al}^{\alpha_1}\delta_{\alpha_2}^{\alpha_3}\delta_{\gamma_1}^{
\gamma_2 }\delta_{\beta_3}^{\beta}\,\al'\,\mathscr{R}^{\star}( R
;{\hmi},\hpl,\hpl) \ ,\label{rq1}
\end{equation}
where:
\begin{flalign}
\mathscr{R}^{\star}( \tmi
;{\hmi},\hpl,\hpl) &\eq R_{st}\,\langle p1\rangle^{2}[q2][13]
~+~\widetilde{R}_{st}\,\langle
p1\rangle^{2}[3q][21]\ ,\nonumber\\[1mm]
\mathscr{R}^{\star}( \hmi
;{\hmi},\hpl,\hpl)&
\eq \frac{1}{\sqrt{3}}R_{st}\,\langle p1\rangle[13]\Big\{2\langle
q1\rangle[q2]-\langle
p1\rangle[p2]\Big\}\nonumber\\ &\qquad\qquad +
~\frac{1}{\sqrt{3}}\widetilde{R}_{st}\,\langle
p1\rangle[12]\Big\{2\langle q1\rangle[q3]-\langle
p1\rangle[p3]\Big\}\ ,\nonumber\\[1mm]
\mathscr{R}^{\star}( \hpl
;{\hmi},\hpl,\hpl) &
\eq \frac{1}{\sqrt{3}}R_{st}\,\langle q1\rangle[13]\Big\{2\langle
p1\rangle[p2]-\langle
q1\rangle[q2]\Big\}
\label{rqstar}\\ &\qquad\qquad + ~\frac{1}{\sqrt{3}}\widetilde{R}_{st}\,\langle
q1\rangle[12]\Big\{2\langle p1\rangle[p3]-\langle
q1\rangle[q3]\Big\}\ ,\nonumber\\[1mm]
\mathscr{R}^{\star}( \tpl
;{\hmi},\hpl,\hpl) &
\eq R_{st}\,\langle q1\rangle^{2}[p2][13]~+ ~\widetilde{R}_{st}\,\langle
q1\rangle^{2}[3p][21]\ .\nonumber\end{flalign}

\subsection{Massive spin 1/2 quark $Q(J=1/2)$}

\noindent
The amplitude describing the decay of $Q(J=1/2)$ into one quark and a gluon is
given by
\begin{eqnarray}
Q(U;\epsilon_1,u_2)&=&(T^{a_1})_{\alpha}^{\al_2}\, g\,\al'
(k_{1\mu}\epsilon_{1\nu}-k_{1\nu}\epsilon_{1\mu})k_3^{\nu}\,
u_2^{\lambda}\sigma^{\mu}_{\lambda\dot
\rho}\bar\chi^{\dot\rho}\nonumber\\[1mm]
&\equiv&(T^{a_1})_{\alpha}^{\al_2}\sqrt{2}g\,{\al'}
\mathscr{Q}(U;\epsilon_1,u_2)\ ,
\end{eqnarray}
where $\bar\chi$ is the lower component of the spinor $U$,
see Eqs. (\ref{u0}) and (\ref{u1}).
The selection rule
\begin{equation}
\mathscr{Q}(U;-,\hpl)=0
\end{equation}
is complementary to Eq. (\ref{selq}) of its higher spin partner
$Q^{\star}(J=3/2)$.
The non-vanishing amplitudes are:
\begin{flalign}
\mathscr{Q}(\hmi;+,\hpl)&
=\langle p2\rangle[12]^2\ ,\\
\mathscr{Q}(\hpl;+,\hpl)&
=\langle q2\rangle[12]^2\ .
\end{flalign}

The amplitude with one quark and two gluons in the final state is given by a
lengthy expression similar to Eq. (\ref{4ptg}), however, as usual, it simplifies
in the helicity basis.
It is convenient to write it as:
\begin{equation}
Q(U;\epsilon_1,\epsilon_2,u_3) ~=~ g^2\,(\al')^{-1}
[ \, V_t \, (T^{a_1}  T^{a_2})^{\al_3} _{\al_4} \
- \ V_u \, (T^{a_2}  T^{a_1})_{\al_4}^{\al_3}\, ]\,
\mathscr{Q}(U;\epsilon_1,\epsilon_2,u_3)\ .
\end{equation}
In this case, the selection rule complementary to (\ref{sel2}) is
\begin{equation}
 \mathscr{Q}(U;-,-,\hpl)=0\ .
\end{equation}
{}For two gluons in the $(+,+)$ helicity configuration, the amplitude reads:
\begin{flalign}
\mathscr{Q}(\hmi;+,+,\hpl)&
=\frac{\langle p3\rangle}{\langle 12\rangle\langle
23\rangle\langle 31\rangle}\ ,
\notag\\
\mathscr{Q}(\hpl;+,+,\hpl)&
=\frac{\langle q3\rangle}{\langle 12\rangle\langle
23\rangle\langle 31\rangle}
\ .\end{flalign}
When the gluons carry opposite helicities, then
\begin{flalign}
\mathscr{Q}(\hmi;+,-,\hpl)&
=~{\al'}^{3/2}\;\frac{[q1][13]^{2}}{[12][23]}\ ,
\notag\\
\mathscr{Q}(\hpl;+,-,\hpl)&
=-{\al'}^{3/2}\frac{[p1][13]^{2}}{[12][23]}\ ,
\end{flalign}
and a similar expression with $(1\leftrightarrow 2)$ for gluons with flipped
helicities.

The sums of the squared moduli of the amplitudes read:
\begin{eqnarray}
\sum_{U=\hmi}^{\hpl}|Q(U;+,+,\hpl)|^2 &=& g^4\,
|\, V_t \, (T^{a_1}  T^{a_2})^{\al_3} _{\al_4}
-  V_u \, (T^{a_2}  T^{a_1})_{\al_4}^{\al_3}\, |^2\frac{1-\al' s}{{\al'}^3
stu}\ ,\nonumber\\
\sum_{U=\hmi}^{\hpl}|Q(U;+,-,\hpl)|^2 &=& g^4\,
|\, V_t \, (T^{a_1}  T^{a_2})^{\al_3} _{\al_4}
-  V_u \, (T^{a_2}  T^{a_1})_{\al_4}^{\al_3}\, |^2\frac{t^2(1-\al' u)}{su}\ , \\
\sum_{U=\hmi}^{\hpl}|Q(U;-,+,\hpl)|^2 &=& g^4\,
|\, V_t \, (T^{a_1}  T^{a_2})^{\al_3} _{\al_4}
-  V_u \, (T^{a_2}  T^{a_1})_{\al_4}^{\al_3}\, |^2\frac{u^2(1-\al' t)}{st}\
.\nonumber
\end{eqnarray}

The amplitudes describing $Q$-decays into two quarks and one antiquark are
described by formulas
similar to (\ref{qq1}), (\ref{qqstar}) in the two-stack case and (\ref{rq1}),
(\ref{rqstar}) in the three-stack case. All what one has to do in order to
obtain the corresponding amplitudes is to replace Eqs. (\ref{qqstar})
and (\ref{rqstar}) by
\begin{flalign}
\mathscr{Q}( \hmi
;{\hpl},\hmi,\hpl) &\eq Q_{su}\,\langle p1\rangle\langle 23\rangle [13]^2
~+~\widetilde{Q}_{su}\,\langle p3\rangle\langle 21\rangle [13]^2\ ,
\nonumber\\[1mm]
\mathscr{Q}( \hpl
;{\hpl},\hmi,\hpl) &
\eq Q_{su}\,\langle q1\rangle\langle 23\rangle [13]^2
~+~\widetilde{Q}_{su}\,\langle q3\rangle\langle 21\rangle [13]^2\ ,
\end{flalign}
and
\begin{flalign}
\mathscr{R}( \hmi
;{\hmi},\hpl,\hpl) &\eq R_{st}\,\langle p2\rangle\langle 13\rangle [23]^2
~+~\widetilde{R}_{su}\,\langle p3\rangle\langle 12\rangle [23]^2\ ,
\nonumber\\[1mm]
\mathscr{R}( \hpl
;{\hmi},\hpl,\hpl) &
\eq R_{st}\,\langle q2\rangle\langle 13\rangle [23]^2
~+~\widetilde{R}_{su}\,\langle q3\rangle\langle 12\rangle [23]^2 \ ,
\end{flalign}
respectively.

\section{Cross sections for the direct production}
After discussing the amplitudes (and their squared moduli) involving one lowest
Regge excitation ($R$, mass $M=1/\al'$) and three massless partons $(p=g,q,\bar
q)$, we collect the results for the subprocesses $p_1(k_1)p_2(k_2)\to p_3(k_3)
R(k_4)$ relevant to the production of
Regge resonances at the LHC. {}For the applications to jet-associated Regge
production, we square the moduli of the amplitudes,
average over helicities and colors of the incident partons and sum over spin
directions (helicity of $p_3$ and $J_z$ of $R$) and colors of the outgoing
particles. In all these processes, quark flavor is a spectator.

The kinematic Mandelstam variables $s,t$ and $u$ have been defined in
Eq. (\ref{mandel}) in such a way that after reverting to the conventional
$(+--\,-)$ metric signature, and crossing to the physical (outgoing) momenta,
$k_3\to -k_3,~ k_4\to -k_4$, they become
\begin{equation}\label{mandel1}
 s=(k_1+k_2)^2\ ,\qquad  t=(k_1-k_3)^2\ ,\qquad u=(k_1-k_4)^2\ ,\end{equation}
satisfying the constraint
\begin{equation}
 s+t+u=M^2
 \end{equation}
due to the momentum conservation $k_1+k_2=k_3+k_4$ and on-shell conditions
$k_1^2=k_2^2=k_3^2=0$, $k_4^2=M^2$.
Their physical domain is
\begin{equation}
s \ > M^2~,\qquad t\ <\ 0~,\qquad u\ <\ 0\ .
\end{equation}

There are some subtleties encountered when analyzing the flow of gauge charges
in the scattering amplitudes, related to the presence of massless and massive
intermediate states expected either to acquire masses due to quantum effects or
to be eliminated by electro-weak symmetry breaking. {}For example, quark-quark
elastic scattering processes
involve exchanges of massless abelian (``color singlet'') gauge bosons
associated to the $U(1)$ ``baryon number'' subgroup of $U(N)$
\cite{Lust:2008qc}. However, it is well-known that the $U(1)$ anomaly generates
their masses  at the one loop level, and certainly affects whole Regge
trajectory. Other processes, like multi-gluon scattering can be also affected by
mass shifts on such a deformed Regge trajectory. In the processes involving
external Regge excitations, this problem becomes even more pronounced because
massless color singlets contribute
to all processes with one or more external quark-antiquark pairs. As an example,
consider the $B(J=2)$ decay into one gluon and one quark-antiquark pair,
described by Eqs. (\ref{bqq1}) and (\ref{bqq2}). Let us focus on the prefactor
\begin{equation}
V_t(T^{a_3}T^{a_4})^{\alpha_1}_{\alpha_2}+
 V_u(T^{a_4}T^{a_3})^{\alpha_1}_{\alpha_2}=\big[2d^{a_3a_4a_n}(V_t+V_u)
+\frac{i}{2}f^{a_3a_4a_n}(V_t-V_u)\big]
(T^{a_n})^{\alpha_1}_{\alpha_2}
\end{equation}
which multiplies the function $\mathscr{B}\sim \langle 12\rangle^{-1}\langle
23\rangle^{-1}\langle 31\rangle^{-1}$. Now consider the limit $\langle
12\rangle\to 0~ (s\to 0$, allowed in the decay channel).
Since $V_t=V_u=1$ in this limit, the amplitude exhibits a massless pole $\langle
12\rangle^{-1}$,
with the residue $\sim d^{a_3a_4a_n}(T^{a_n})^{\alpha_1}_{\alpha_2}$. The pole
is due to
intermediate gauge bosons, produced in the $B$ decay together with one free
gluon [see Eq. (\ref{bamp2})], and subsequently decaying into the
quark-antiquark
pair. Note that the $U(1)$ generator $(T^{a_n}=Q_A I\!\! I_N, ~Q_A=1/\sqrt{2N})$
is among these gauge bosons and there is no obvious way to remove it
from the disk amplitude. A formal $N\to \infty$ limit would help in that respect
by suppressing such singlet contributions. When collecting the squared
amplitudes describing direct production of Regge resonances,
we  set the number of colors to $N=3$, but we  display the abelian
coupling $Q_A=1/\sqrt{6}$ explicitly.
We always assume that the external partons are either color octet gluons or
color triplet quarks (or antitriplet antiquarks), however we allow the
possibility of color singlet Regge excitations $B_0$ and $\Phi_0$ labeled by an
additional subscript $0$.

The following formulas, valid for general $N$, are useful for summing over the
non-abelian color indices:
\begin{flalign}
\sum_{a_1,a_2,a_3}d^{a_1a_2a_3}d^{a_1a_2a_3}&  ={(N^2-1)(N^2-4)\over 16
N}~,\qquad
\big(d^{a_1a_20} =\frac{Q_A}{2}\delta^{a_1a_2}\big) \cr
\sum_{a_1,a_2}f^{i_1a_1a_2}f^{i_2a_1a_2}&  = N\
\delta^{i_1i_2}~,\qquad\qquad\qquad\quad \big(f^{a_1a_20}=0\big)\cr
\sum_{a_1,a_2,a_3,a_4}t^{a_1a_2a_3a_4}t^{a_1a_2a_3a_4}&  =-{(N^2-1)(N^2-4)\over
8}\cr
\sum_{a_1,a_2,a_3,a_4}t^{a_1a_2a_3a_4}t^{a_2a_3a_1a_4}&  =0
\end{flalign}

In the Tables below, we collect the squared amplitudes for all disk-level
production mechanisms of Regge resonances, listed in order of the initial
two-particle channels: $gg$ followed by $gq$ and $q\bar q$. The quark-quark
channel can be obtained from $q\bar q$ by trivial crossing.
Except for the case of four-fermion processes, we factored out the QCD coupling
factor $g^4$.
\newpage
{\centerline{\noindent{\bf Table 1: }{\it Gluon fusion}}
\vbox{{$$
\vbox{\offinterlineskip\tabskip=0pt
\halign{\strut\vrule#
&~$#$~\hfil
&\vrule$#$
&~$#$~\hfil
&\vrule$#$\cr
\noalign{\hrule}
& && &\cr
&{\rm subprocess}  &&\hskip4.5cm|\Mc|^2/g^4 & \cr
& && &\cr
\noalign{\hrule}\noalign{\hrule}
& && &\cr
& gg\to gB &&
\textstyle \frac{5}{8}\!\left(\, V_s^2+ V_t^2
+V_u^2\,\right)\frac{(s-M^2)^4+(t-M^2)^4+(u-M^2)^4}{M^2stu}
       &\cr
& && &\cr
\noalign{\hrule}
& && &\cr
&  gg\to gB_0 &&
\textstyle {3\over 4}Q_A^2\!\left(\, V_s+ V_t
+V_u\,\right)^2 \frac{(s-M^2)^4+(t-M^2)^4+(u-M^2)^4}{M^2stu}
       &\cr
& && &\cr
\noalign{\hrule}
& && &\cr
&gg\to g\Phi   &&
\textstyle {5\over 8}\!\left(\, V^2_s+ V^2_t
+V^2_u\,\right) \frac{s^4+t^4+u^4+M^8}{M^2stu}
    &\cr
& && &\cr
\noalign{\hrule}
& && &\cr
&gg\to g\Phi_0&&
\textstyle {3\over 4}Q_A^2\!\left(\, V_s+ V_t
+V_u\,\right)^2 \frac{s^4+t^4+u^4+M^8}{M^2stu}
    &\cr
& && &\cr
\noalign{\hrule}
& && &\cr
&gg\to \bar qQ&&
\textstyle \frac{1}{4}\big[
\frac{3}{32}(V_t+V_u)^2+(\frac{5}{96}
+\frac{Q_A^2}{8})(V_t-V_u)^2\big]
\frac{(s-M^2)M^6+(t-M^2)u^3+(u-M^2)t^3}{M^2stu}
    &\cr
& && &\cr
\noalign{\hrule}
& && &\cr
&gg\to\bar qQ^{\star} &&\textstyle \big[
\frac{3}{32}(V_t+V_u)^2+(\frac{5}{96}
+\frac{Q_A^2}{8})(V_t-V_u)^2\big]
\frac{(s-M^2)^3M^2+(t-M^2)^3u+(u-M^2)^3t}{M^2stu}
    &\cr
& && &\cr
\noalign{\hrule}}}$$
\vskip-6pt
\vskip10pt}}}
\noindent
\newpage
{\centerline{\noindent{\bf Table 2: }{\it Quark-gluon absorption}}
\vbox{{$$
\vbox{\offinterlineskip\tabskip=0pt
\halign{\strut\vrule#
&~$#$~\hfil
&\vrule$#$
&~$#$~\hfil
&\vrule$#$\cr
\noalign{\hrule}
& && &\cr
&{\rm subprocess}  &&\hskip4.5cm|\Mc|^2/g^4 & \cr
& && &\cr
\noalign{\hrule}\noalign{\hrule}
& && &\cr
& qg\to qB &&
\textstyle -\frac{1}{16}\big[
(V_s-V_u)^2+(\frac{5}{9}
+\frac{4Q_A^2}{3})(V_s+V_u)^2\big]
\frac{[(s-M^2)^2 +(u-M^2)^2](tM^2+4su)}{M^2stu}
       &\cr
& && &\cr
\noalign{\hrule}
& && &\cr
&  gq\to qB_0 &&
\textstyle -\frac{Q_A^2}{12}(V_s+V_u)^2
\frac{[(s-M^2)^2 +(u-M^2)^2](tM^2+4su)  }{M^2stu}
       &\cr
& && &\cr
\noalign{\hrule}
& && &\cr
&qg\to q\Phi   &&
\textstyle -\frac{1}{4}\big[
(V_s-V_u)^2+(\frac{5}{9}
+\frac{4Q_A^2}{3})(V_s+V_u)^2\big]
\frac{s^2 +u^2}{M^2t}
    &\cr
& && &\cr
\noalign{\hrule}
& && &\cr
&qg\to q\Phi_0&&
\textstyle  -
\frac{Q_A^2}{3}(V_s+V_u)^2
\frac{s^2 +u^2}{M^2t}
    &\cr
& && &\cr
\noalign{\hrule}
& && &\cr
&qg\to qW   &&
\textstyle -\frac{3}{16}\big[
(V_s-V_u)^2+(\frac{5}{9}
+\frac{4Q_A^2}{3})(V_s+V_u)^2\big]
\frac{(s-M^2)^2 +(u-M^2)^2}{su}
    &\cr
& && &\cr
\noalign{\hrule}
& && &\cr
&qg\to qW_0&&
\textstyle  -
\frac{Q_A^2}{4}(V_s+V_u)^2
\frac{(s-M^2)^2 +(u-M^2)^2}{su}
    &\cr
& && &\cr
\noalign{\hrule}
& && &\cr
&gq\to  gQ&&
\textstyle -\frac{1}{16}\big[
(V_s+V_u)^2+(\frac{5}{9}
+\frac{4Q_A^2}{3})(V_s-V_u)^2\big]
\frac{(t-M^2)^3M^2+(u-M^2)^3s+(s-M^2)^3u}{M^2stu}
    &\cr
& && &\cr
\noalign{\hrule}
& && &\cr
&gq\to gQ^{\star} &&
\textstyle -\frac{1}{4}\big[
(V_s+V_u)^2+(\frac{5}{9}
+\frac{4Q_A^2}{3})(V_s-V_u)^2\big]
\frac{(t-M^2)M^6+(u-M^2)s^3+(s-M^2)u^3}{M^2stu}
    &\cr
& && &\cr
\noalign{\hrule}}}$$
\vskip-6pt
\vskip10pt}}}
\noindent
\newpage
{\centerline{\noindent{\bf Table 3: }{\it Quark-antiquark annihilation}}
\vbox{{$$
\vbox{\offinterlineskip\tabskip=0pt
\halign{\strut\vrule#
&~$#$~\hfil
&\vrule$#$
&~$#$~\hfil
&\vrule$#$\cr
\noalign{\hrule}
& && &\cr
&{\rm subprocess}  &&\hskip4.5cm|\Mc|^2/g^4 & \cr
& && &\cr
\noalign{\hrule}\noalign{\hrule}
& && &\cr
&\bar q q\to gX &&
\textstyle |{\cal M}(\bar q q\to gX)|^2=-\frac{8}{3}|{\cal M}(qg\to
qX)|^2(s\to  u,~ u\to t,~ t\to s)
       &\cr
&X=B,\Phi,W &&
       &\cr
\noalign{\hrule}
& && &\cr
&   &&
\Big\{\! -\frac{t^2}{4}(su|Q_{su}+\widetilde{Q}_{su}|^2+ut|Q_{su}|^2
+st|\widetilde{Q}_{su}|^2 )
    &\cr
&\bar q q\to \bar q Q  &&\quad
+\,\frac{t^2}{12}\,\big[su(Q_{su}+\widetilde{Q}_{su})(Q_{us}+\widetilde{Q}_{us}
)^*
+utQ_{su}\widetilde{Q}_{us}^*
+st\widetilde{Q}_{su}{Q}_{us}^*\big]
 &\cr
& &&\quad -\,\frac{u^2}{4}(st|R_{st}+\widetilde{R}_{st}|^2 +ut|R_{st}|^2
+su|\widetilde{R}_{st}|^2\Big\} ~+~ \Big\{ s\leftrightarrow u\Big\}&\cr
& && &\cr
\noalign{\hrule}
& && &\cr
& \bar q q\to \bar q' Q'  &&
-\frac{t^2}{4}(su|Q_{su}+\widetilde{Q}_{su}|^2+ut|Q_{su}|^2
+st|\widetilde{Q}_{su}|^2) ~+~ \Big(Q\to R  ~;~ u\leftrightarrow t\Big)
    &\cr
& && &\cr
\noalign{\hrule}
& && &\cr
&   &&
\Big\{\! -\frac{(M^2-t)^2}{4}(su|Q_{su}+\widetilde{Q}_{su}|^2+ut|Q_{su}|^2
+st|\widetilde{Q}_{su}|^2 )+\frac{t(M^2-t)}{6}|uQ_{su}-s\widetilde{Q}_{su}|^2
    &\cr
&   &&
\quad +\,\frac{(M^2-t)^2}{12}\big[su(Q_{su}+\widetilde{Q}_{su})(Q_{us}+
\widetilde{Q}_{us})^*+utQ_{su}\widetilde{Q}_{us}^*
+st\widetilde{Q}_{su}Q_{us}^* \big]
    &\cr
& \bar q q\to \bar q Q^{\star}   &&
\quad
+\,\frac{t(M^2-t)}{18}(uQ_{su}-s\widetilde{Q}_{su})(u\widetilde{Q}_{us}-sQ_{us}
)^*
    &\cr
& &&\quad
-\frac{(M^2-u)^2}{4}(st|R_{st}+\widetilde{R}_{st}|^2+ut|R_{st}|^2
+su|\widetilde{R}_{st}|^2
)+\frac{u(M^2-u)}{6}|tR_{st}-s\widetilde{R}_{st}|^2\Big\}
&\cr
& &&\qquad\qquad\qquad\qquad\qquad\qquad +~ \Big\{ s\leftrightarrow u\Big\}&\cr
& && &\cr
\noalign{\hrule}
& && &\cr
& \bar q q\to \bar q' {Q'}^{\star}  &&
 -\frac{(M^2-t)^2}{4}(su|Q_{su}+\widetilde{Q}_{su}|^2+ut|Q_{su}|^2
+st|\widetilde{Q}_{su}|^2 )+\frac{t(M^2-t)}{6}|uQ_{su}-s\widetilde{Q}_{su}|^2
    &\cr
&     &&\qquad\qquad\qquad\qquad\quad ~~+~ \Big(Q\to R  ~;~ u\leftrightarrow
t\Big)
    &\cr
    & && &\cr
\noalign{\hrule}
}}$$
\vskip-6pt
\vskip10pt}}}
\noindent

\break
\begin{acknowledgments}
We are grateful to Ignatios Antoniadis, Massimo Bianchi and Daniel H\"artl
for useful discussions.
This work  is supported in part by the European Commission under Project MRTN-CT-2004-005104.
The research of T.R.T.\ is supported by the U.S.  National Science
Foundation Grants PHY-0600304 and PHY-0757959.
He is grateful to the Center for Advanced Studies at Ludwig--Maximilians--Universit\"at
M\"unchen, for its kind hospitality.
Any opinions, findings, and conclusions or
recommendations expressed in this material are those of the authors and do not necessarily
reflect the views of the National Science Foundation.
\end{acknowledgments}

\nocite{*}
\bibliography{lhc}
\end{document}